\documentclass{aa}
\usepackage{graphicx}
\usepackage{txfonts}

\begin{document}

\title{N$_2$H$^+$(1--0) survey of massive molecular cloud cores}

\author{L. Pirogov \inst{1}, I. Zinchenko \inst{1,2,3},
P. Caselli \inst{4}, L. E. B. Johansson \inst{5}, P. C. Myers \inst{6}}

\offprints{L. Pirogov}

\institute{
Institute of Applied Physics of the Russian Academy of Sciences, Ulyanova 46,
603950 Nizhny Novgorod, Russia \\
\email{pirogov@appl.sci-nnov.ru} \and
Nizhny Novgorod University , Gagarin av. 23,
603950 Nizhny Novgorod, Russia \and
Helsinki University Observatory, T\"ahtitorninm\"aki, P.O. Box 14, FIN-00014
University of Helsinki, Finland \and
INAF -- Osservatorio Astrofisico di Arcetri, Largo E. Fermi 5, I-50125 Firenze,
Italy \and
Onsala Space Observatory, S-43992, Onsala, Sweden \and
Harvard-Smithsonian Center of Astrophysics, 60 Garden Street, Cambridge,
MA 02138, USA}

\date{}

\abstract{
We present the results of N$_2$H$^+$(1--0) observations of
35 dense molecular cloud cores from the northern and southern hemispheres
where massive stars and star clusters are formed.
Line emission has been detected in 33 sources, for 28 sources
detailed maps have been obtained.
Peak N$_2$H$^+$ column densities lie in the range:
$3.6\cdot 10^{12}-1.5\cdot 10^{14}$~cm$^{-2}$.
Intensity ratios of (01--12) to (23--12) hyperfine components are
slightly higher than the LTE value.
The optical depth of (23--12) component toward peak intensity positions
of 10 sources is $\sim 0.2-1$.
In many cases the cores have
elongated or more complex structures with several emission peaks.
In total, 47 clumps have been revealed in 26 sources.
Their sizes lie in the range 0.3--2.1~pc, the range of
virial masses is $\sim 30-3000~M_{\odot}$.
Mean N$_2$H$^+$ abundance for 36 clumps is $5\cdot 10^{-10}$.
Integrated intensity maps with aspect ratios $<2$
have been fitted with a power-law radial distribution $\sim r^{-p}$
convolved with the telescope beam.
Mean power-law index for 25 clumps is close to 1.3.
For reduced maps where positions of low intensity are rejected
mean power-law index is close to unity corresponding to the $\sim r^{-2}$
density profile provided N$_2$H$^+$ excitation conditions do not vary
inside these regions.
In those cases where we have relatively extensive and high quality maps,
line widths of the cores either decrease or stay constant with distance
from the center, implying an enhanced dynamical activity in the center.
There is a correlation between total velocity gradient direction
and elongation angle of the cores.
However, the ratio of rotational to gravitational energy
is too low (4$\cdot$10$^{-4}$--7.1$\cdot$10$^{-2}$)
for rotation to play a significant role in the dynamics of the cores.
A correlation between mean line widths and sizes of clumps has been found.
A comparison with physical parameters of low-mass cores is given.
}

\authorrunning{L. Pirogov et al.}

\titlerunning{N$_2$H$^+$(1--0) survey of massive cores}

\maketitle

\section{Introduction}
\label{sec:intro}

It is well-known that stars in some cases are formed in clusters
and isolation in others, however, the reasons of such a dichotomy
are far from being understood at present.
The number of stars depends apparently on physical
parameters of dense molecular cloud cores where stars are formed.
It is known, in particular, that clusters
are formed in the most massive and turbulent cores.
High-mass stars are encountered solely in such cores.
An important feature of massive cores
is their clumpy structure which extends from large scales down
to the smallest ones even unresolved by present-day telescopes.
An existence of small-scale clumpiness is implied,
for example, by 1--2 orders of magnitude difference between mean densities
and densities derived from multitransitional analysis (e.g. Bergin et al.
\cite{bergin1}, Zinchenko \cite{zin2}, Zinchenko et al. \cite{zin4});
by the absence of significant broadening of presumably very optically
thick CS lines as compared to the rarer C$^{34}$S isotopomer lines
(Zinchenko et al. \cite{zin4}); by the hyperfine intensity ratios anomalies
in HCN (Pirogov \cite{pirogov}) and ammonia
(Stutzki \& Winnewisser \cite{stutzki}) in warm clouds.
The degree of small-scale clumpiness is probably related
to the number of newborn stars.

Important parameters for describing evolution of star forming cores
in modern theoretical models are radial dependences of density
and velocity dispersion in envelopes that surround newborn stars.
A popular model for massive star formation is based on the logatropic
equation of state which incorporates non-thermal gas motions
(McLaughlin \& Pudritz \cite{mclaughlin}).
It results in the $\sim r^{-1}$ density distribution.
Velocity dispersion in this model is nearly constant in the center
and rises with distance in the outer regions as $\sim r^{1/3}$.
In contrast, the standard theory of low-mass star formation
from a singular isothermal sphere (Shu \cite{shu}) predicts
the $\sim r^{-2}$ density profile in the static outer envelope.
Both models predict $\sim r^{-3/2}$ and $\sim r^{-1/2}$ density
and velocity profiles, respectively, for the regions involved in
free-falling collapse (Shu \cite{shu}, McLaughlin \& Pudritz \cite{mclaughlin1}).
Therefore, studies of density and velocity dispersion profiles in the cores
are essential for discrimination between different theoretical models
and for determination of their dynamical status.

Recently, density structure of large samples of objects have been explored
in several studies.
Analysis of dust continuum emission in low-mass cores
where stars are formed mostly in isolation, reveals a density power-law index of
$\alpha\sim 1.5-2$ (e.g. Motte \& Andr\'e \cite{motte}, Shirley et al. \cite{shirley})
being more or less in accordance with the isothermal sphere model.
Dust continuum emission studies in high-mass star forming regions
have resulted in the following power-law indices:
$\langle\alpha\rangle=1.6(0.5)$ (Beuther et al. \cite{beuther}) and
$\langle\alpha\rangle=1.8(0.4)$ (Mueller et al. \cite{mueller}), while
molecular multi-line modeling of several massive star forming regions
(Van der Tak et al. \cite{tak}) gave somewhat lower values: $\alpha=1.0-1.5$.

In their study of velocity dispersion profiles in low-mass cores,
Goodman et al. (\cite{goodman98}) found nearly constant velocity dispersion
within so-called ``coherent regions" with sizes of $\sim 0.1$~pc and
a rise of velocity dispersion in the outer regions with power-law index
$\ga 0.2$, which is more consistent with the logatropic model.
In contrast, molecular line studies of high-mass cores
show highly non-thermal lines of decreasing widths with distance
from center (Zinchenko \cite{zin2}, Lapinov et al. \cite{lapinov},
Fontani et al. \cite{fontani}), which apart from optical depth effects
could imply a higher degree of gas dynamical activity
in central regions.

Recently, about 60 nearby low-mass cores were mapped in the N$_2$H$^+$(1--0)
line (Caselli et al. \cite{paper1}, hereafter Paper~I).
As a result, a number of physical parameters including
density and velocity dispersion profiles were derived.
N$_2$H$^+$ was chosen as a tracer of quiescent high density gas (critical
density for excitation of the $J=1-0$ transition
at a kinetic temperature of 40~K is $4.4\cdot 10^5$~cm$^{-3}$).
An advantage of the N$_2$H$^+$(1--0) transition is the hyperfine structure
which enables a reliable determination of the optical depth and improves
estimates of related physical parameters of the dense gas that surrounds
young stars.
It is known that the N$_2$H$^+$(1--0) lines have relatively strong intensities
and moderate ($\tau\sim 1$) optical depths both in cold and warm clouds
(Womack et al. \cite{womack}) associated with low-mass and high-mass
star forming regions, respectively.
The goal of the present study is to derive physical parameters
for a representative sample of dense cores in high-mass star forming regions
from observations in the N$_2$H$^+$(1--0) line.
The derived parameters will be compared with those of low-mass cores
(Paper~I) to help to understand the details of the star formation process.

A sample list for the study has been taken from the
ammonia database compiled by P.~Myers and co-workers (Jijina et al. 1999)
where data on cluster existence is included.
This list overlaps with a sample of massive cores earlier mapped in CS,
both in the northern and southern hemispheres
(Zinchenko et al. \cite{zin1}, \cite{zin3}, \cite{zin4}; Juvela \cite{juvela}).
In the present paper we give the results of N$_2$H$^+$(1--0)
observations towards a sample of 35 massive dense cores
where star formation in clusters is going on.
Their physical parameters including density
and velocity dispersion profiles are derived and compared
with parameters of low-mass cores (Paper~I).

\section{Observations}

The N$_2$H$^+$(1--0) observations at the frequency 93.2~GHz have been done
in April, 2000 at 20-m OSO radiotelescope and in February, 2001
at 15-m SEST radiotelescope.
The observing parameters are given in Table~\ref{table:param}.

For OSO we took into account beam efficiency dependence on source elevation
and give an average value.
We also give the ranges of system temperatures which depend
on source elevation and weather conditions.
At OSO we used both a filterbank spectrum analyzer (256 channels)
and autocorrelator (1600 channels) while at SEST an acousto-optical
spectrum analyzer (1000 channels) was used.
Pointing was regularly checked by SiO maser observations and
usually was better than 5$\arcsec$.

\begin{table}[htb]
\centering
\caption[]{N$_2$H$^+$(1--0) observing parameters.}
\begin{tabular}{lccll}\hline\noalign{\smallskip}
Telescope     &
$\Delta \Theta$  &
$\eta_{\rm mb} $ &
$T_{\rm sys}$       &
$\delta V$         \\
         & ($\arcsec$) &           & (K)       & (km s$^{-1}$)\\
\noalign{\smallskip}\hline\noalign{\smallskip}
OSO-20m  & 40     & $\sim 0.5$   & 250--500  & 0.8; 0.16\\
SEST-15m & 55     & 0.75         & 200--300  & 0.14\\
\noalign{\smallskip}\hline\noalign{\smallskip}
\end{tabular}
\label{table:param}
\end{table}

The objects for observations were selected from the list of dense cores
observed earlier in the CS(2--1) line (Zinchenko et al. \cite{zin1},
\cite{zin3}, \cite{zin4}; Juvela \cite{juvela})
as well as from the ammonia database (Jijina et al. \cite{jijina})
according to the following criteria:
an existence of embedded clusters of stars detected in infrared
(Jijina et al. \cite{jijina}, Hodapp \cite{hodapp},
Carpenter et al. \cite{carpenter}),
distances to the objects not far than 5~kpc
and an existence of
extended CS or NH$_3$ emission regions ($>2'$) in order to get
detailed N$_2$H$^+$(1--0) maps.
In the cases of no data on the clusters existence,
the sources with high IR-luminosities were selected
($L>10^4$ L$_{\odot}$), which can be an indirect indication
of cluster (Jijina et al. \cite{jijina}).
In total, 35 objects that satisfy these criteria were selected.
The source list with coordinates and distances is given
in Table~\ref{table:list}.
The number of points per map and the names of associated objects
are also given.
Mapping has been done with 20$\arcsec$ grid spacing (Nyquist sampled maps)
with the exception of S~187, G~133.69 and S~153 for which the spacing
was 40$\arcsec$ (beam sampled maps).

\begin{table*}[htb]
\centering
\caption[]{Source list.}
\begin{tabular}{lrrccl}
\noalign{\hrule}\noalign{\smallskip}
Source
          & RA (1950)                      & Dec (1950)                   & $D$    &Points  & Associated \\
          & (${\rm (^h)\  (^m)\  (^s)\ }$) &($\degr$ $\arcmin$ $\arcsec$) & (kpc)  &per map & objects    \\
\noalign{\smallskip}\hline\noalign{\smallskip}
RNO~1B        &00 33 53.3   &63 12 32    &  0.85$^a$  & 70  & L1287, G~121.30+0.66, IRAS~00338+6312\\
NGC~281       &00 49 29.2   &56 17 36    &  3.5 $^b$  & 34  & S184, G~123.07-6.31 \\
S~187         &01 19 58.0   &61 34 32    &  1.0 $^c$  & 67  & G~126.68-0.83       \\
G~133.69+1.22 &02 21 40.8   &61 53 26    &  2.1 $^d$  & 52  & W3                  \\
S~199         &02 57 36.9   &60 17 32    &  2.1 $^d$  & 46  & IC1848, AFGL~4029   \\
S~201         &02 59 22.4   &60 16 12    &  2.1 $^e$  & 41  & AFGL~437, G~138.50+1.64 \\
AFGL~490      &03 23 34.0   &58 33 47    &  0.9 $^f$  & 18  \\
G142.00+1.83  &03 23 41.0   &58 36 52    &  0.9 $^f$  & 37  & AFGL~490\\
Per~4         &03 26 12.1   &31 17 13    &  0.35$^g$  & 78  \\
LkH$\alpha$101     &04 26 57.3   &35 09 56    &  0.8 $^h$  & 9   & AFGL~585, S222\\
G~173.48+2.45 &05 35 51.3   &35 44 16    &  2.3 $^d$  & 54  & S231 \\
G~173.58+2.44 &05 36 06.3   &35 39 21    &  2.3 $^d$  & 41\\
AFGL~6366     &06 05 40.1   &21 31 16    &  2.0 $^i$  & 50  & S247\\
S~255         &06 09 58.2   &18 00 17    &  2.5 $^d$  & 67  & AFGL~896\\
LkH$\alpha$25      &06 37 59.5   &09 50 53    &  0.8 $^j$  & 1   & AFGL~4519S, NGC~2264, S273\\
S~76E         &18 53 47.0   &07 49 25    &  2.1 $^k$  & 61  & G40.50+2.54\\
S~88B         &19 44 41.4   &25 05 17    &  2.0 $^d$  & 25\\
DR~21         &20 37 13.5   &42 12 15    &  3.0 $^l$  & 56  & W75\\
S~140         &22 17 41.1   &63 03 44    &  0.9 $^d$  & 82  & L1204\\
L~1251T4      &22 27 10.2   &74 58 00    &  0.35$^m$  & 119\\
S~153         &22 56 42.0   &58 28 45    &  4.0 $^d$  & 42\\
G~264.28$+$1.48  &08 54 39.0    &$-$42 53 30   & 1.4$^n$ & 43\\
G~265.14$+$1.45  &08 57 36.3    &$-$43 33 38   & 1.7$^n$ & 64\\
G~267.94$-$1.06  &08 57 21.7    &$-$47 19 04   & 0.7$^n$ & 9\\
G~268.42$-$0.85  &09 00 12.1    &$-$47 32 07   & 1.3$^n$ & 60\\
G~269.11$-$1.12  &09 01 51.6    &$-$48 16 42   & 2.6$^n$ & 53\\
G~270.26$+$0.83  &09 14 58.0    &$-$47 44 00   & 2.6$^n$ & 27\\
G~285.26$-$0.05  &10 29 36.8    &$-$57 46 40   & 4.7$^n$ & 59\\
G 291.27$-$0.71  &11 09 42.0    &$-$61 01 55   & 2.7$^o$ & 104\\
G~294.97$-$1.73  &11 36 51.6    &$-$63 12 09   & 1.2$^n$ & 39\\
G~305.36$+$0.15  &13 09 21.0    &$-$62 21 43   & 4.2$^n$ & 18\\
G~316.77$-$0.02  &14 41 10.4    &$-$59 35 50   & 3.1$^p$ & 136 \\
G~345.01$+$1.80  &16 53 17.3    &$-$40 09 23   & 2.1$^p$ & 103 & RCW~116\\
G~345.41$-$0.94  &17 06 02.2    &$-$41 32 06   & 2.8$^p$ & 71  & RCW~117\\
G~351.41$+$0.64  &17 17 32.3    &$-$35 44 03   & 1.7$^q$ & 88  & NGC~6334I\cr
\noalign{\smallskip}\hline\noalign{\smallskip}
\end{tabular}

$^a$ Yang et al. (\cite{yang}),
$^b$ Martin \& Henning (\cite{martin}),
$^c$ Fich \& Blitz (\cite{fich}),
$^d$ Blitz et al. (\cite{blitz}),
$^e$ assumed,
$^f$ Harvey (\cite{harvey2}),
$^g$ Paper~I,
$^h$ Lada (\cite{lada}),
$^i$ Carpenter et al. (\cite{carpenter}),
$^j$ Snell et al. (\cite{snell}),
$^k$ Plume et al. (\cite{plume}),
$^l$ Harvey et al. (\cite{harvey1}),
$^m$ Toth \& Walmsley (\cite{toth}),
$^n$ Zinchenko et al. (\cite{zin3}),
$^o$ Brand \& Blitz (\cite{brand}),
$^p$ Juvela (\cite{juvela}),
$^q$ Neckel (\cite{neckel})

\label{table:list}
\end{table*}

\normalsize

\section{Results}

\subsection{Integrated intensity maps}

The N$_2$H$^+$(1--0) emission has been detected in 33 of 35 sources
(except LkH$\alpha$101 and LkH$\alpha$25).
Maps of intensities integrated over all hyperfine components
for 31 object are given in Fig.~\ref{maps}.
The maps of AFGL~490 and G~142.00
(which actually belong to a single molecular complex),
S~88B, G~267.94 and G~305.36 have not been completed
(the latter two are not shown in Fig.~\ref{maps}).
The maps of S~187, G~133.69 and AFGL~6366  each consist
of two separated clumps at the 30\% peak intensity level.
Some others (e.g. S~255, G~316.77, G~345.41, G~351.41) contain
intensity peaks separated at the half maximum intensity level.
The maps of RNO~1B, NGC~281, G~173.48, S~76E, G~270.26 indicate
spherical symmetry while others demonstrate elongated, curved
or more complex structures.
Per~4 and L~1251T4 show the most narrow and weak lines and actually
belong to low-mass star forming cores (see Paper~I).
The clumpy structures visible in the maps of these two sources
(see Fig.~\ref{maps}) are primarily caused by noise.
The OSO and SEST beam sizes are shown in Fig.~\ref{maps} on
RNO~1B and G~268.42 maps, respectively.

Most of the sources have associated IRAS point sources.
The positions of IRAS sources are indicated by stars and
the uncertainty ellipses corresponding to 95\% confidence level
are also shown.
In 17 objects IRAS sources
lie within the half maximum level (marked by thick black lines)
or coincide with a secondary intensity peak.
Although there is no IRAS point source in DR~21, it is a well-known
star forming molecular cloud with compact H~II regions
and strong extended far-infrared emission (Harvey et al. \cite{harvey3})
which is close in shape to the N$_2$H$^+$(1--0)
integrated intensity distribution.

\begin{figure}
 \centering \includegraphics[width=9cm]{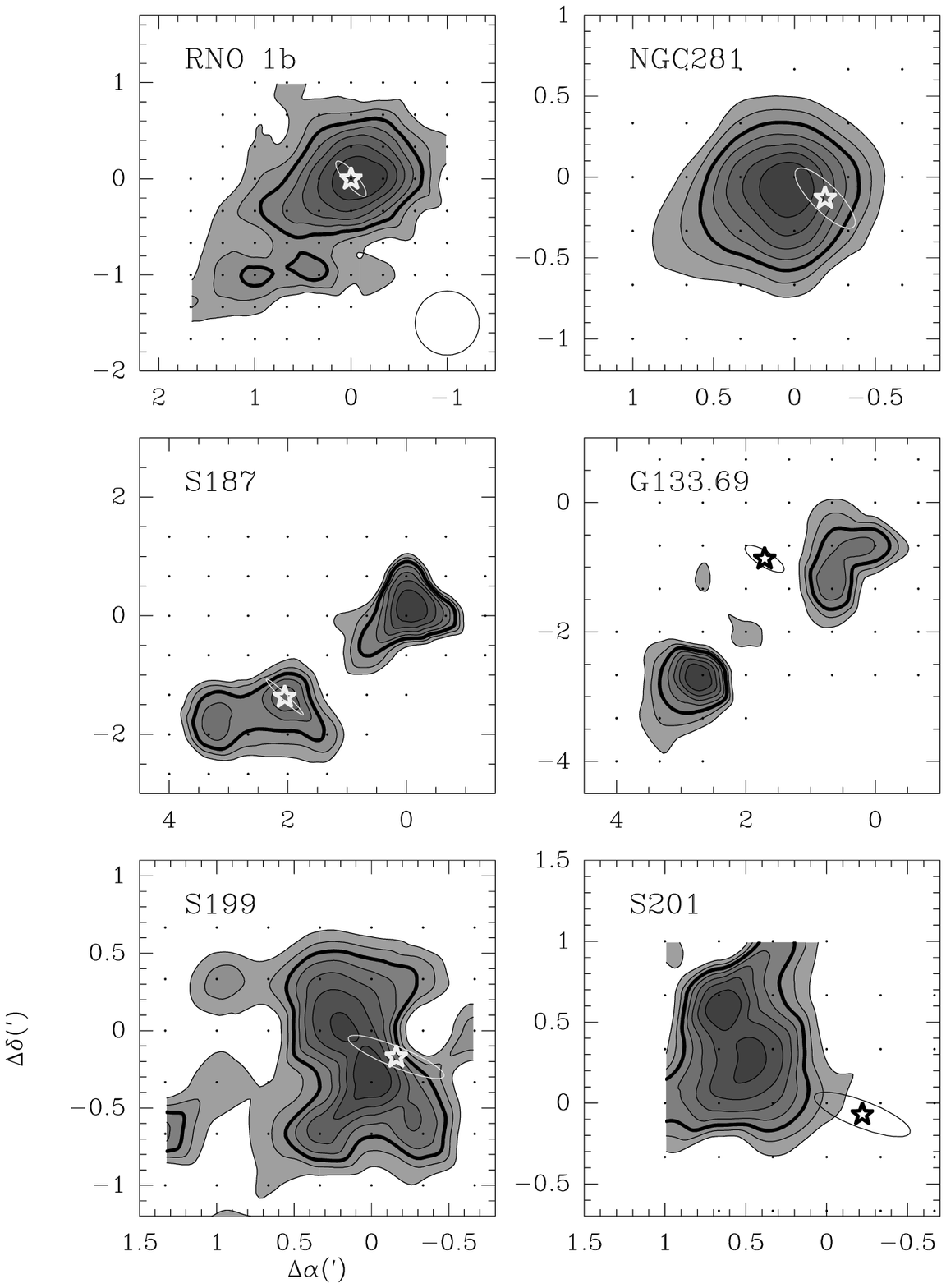}
\caption{N$_2$H$^+$(1--0) integrated intensity maps.
Intensity contour ranges from 30\% to 90\% of peak intensity,
thick contours correspond to half maximum levels.
Observed positions are marked by dots and
IRAS point sources are marked by stars.
The uncertainty ellipses corresponding to 95\% confidence level
in IRAS point source position are also shown}
\label{maps}
\end{figure}

\addtocounter{figure}{-1}

\begin{figure}
 \centering \includegraphics[width=9cm]{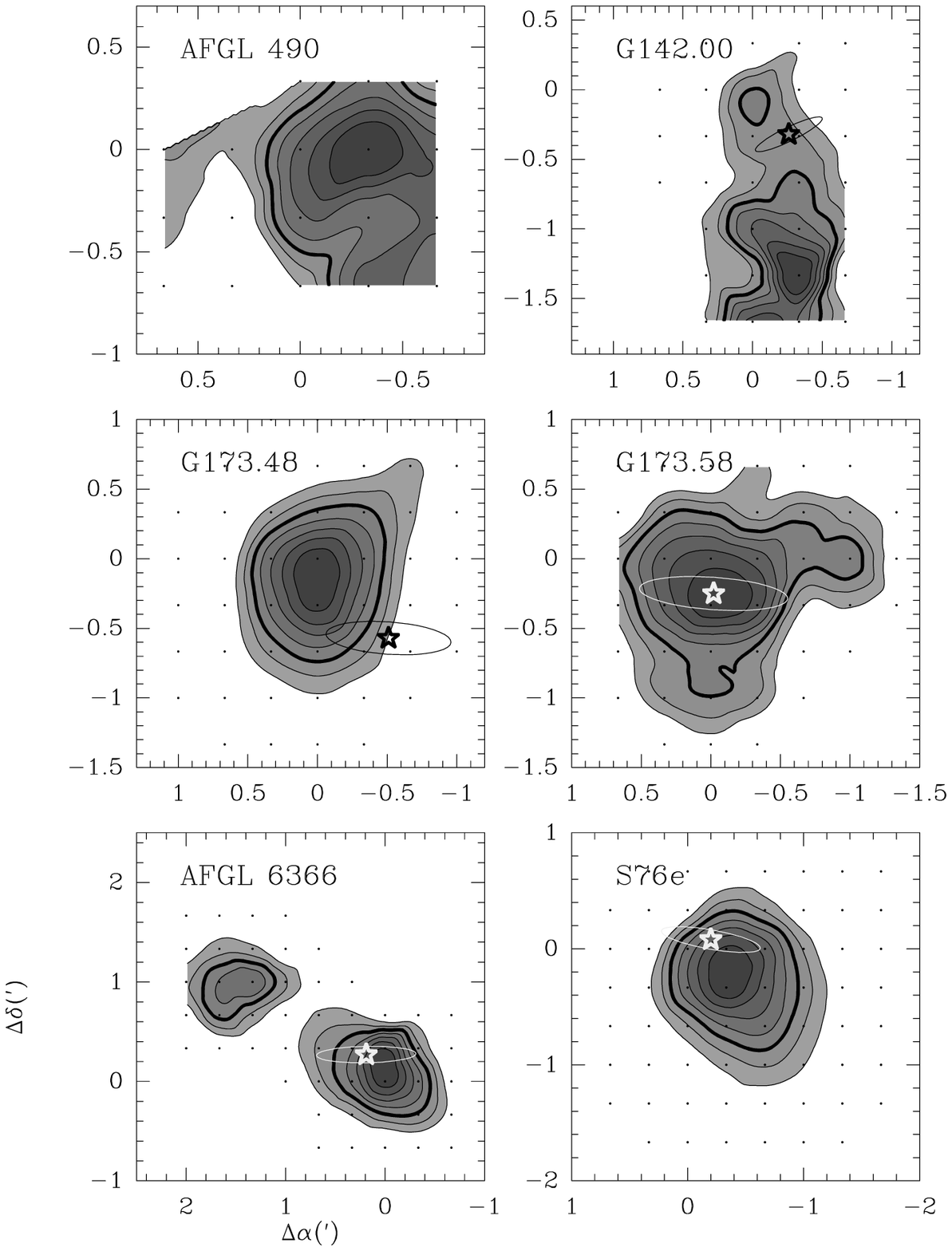}
\caption{continued}
\end{figure}

\addtocounter{figure}{-1}

\begin{figure}
 \centering \includegraphics[width=9cm]{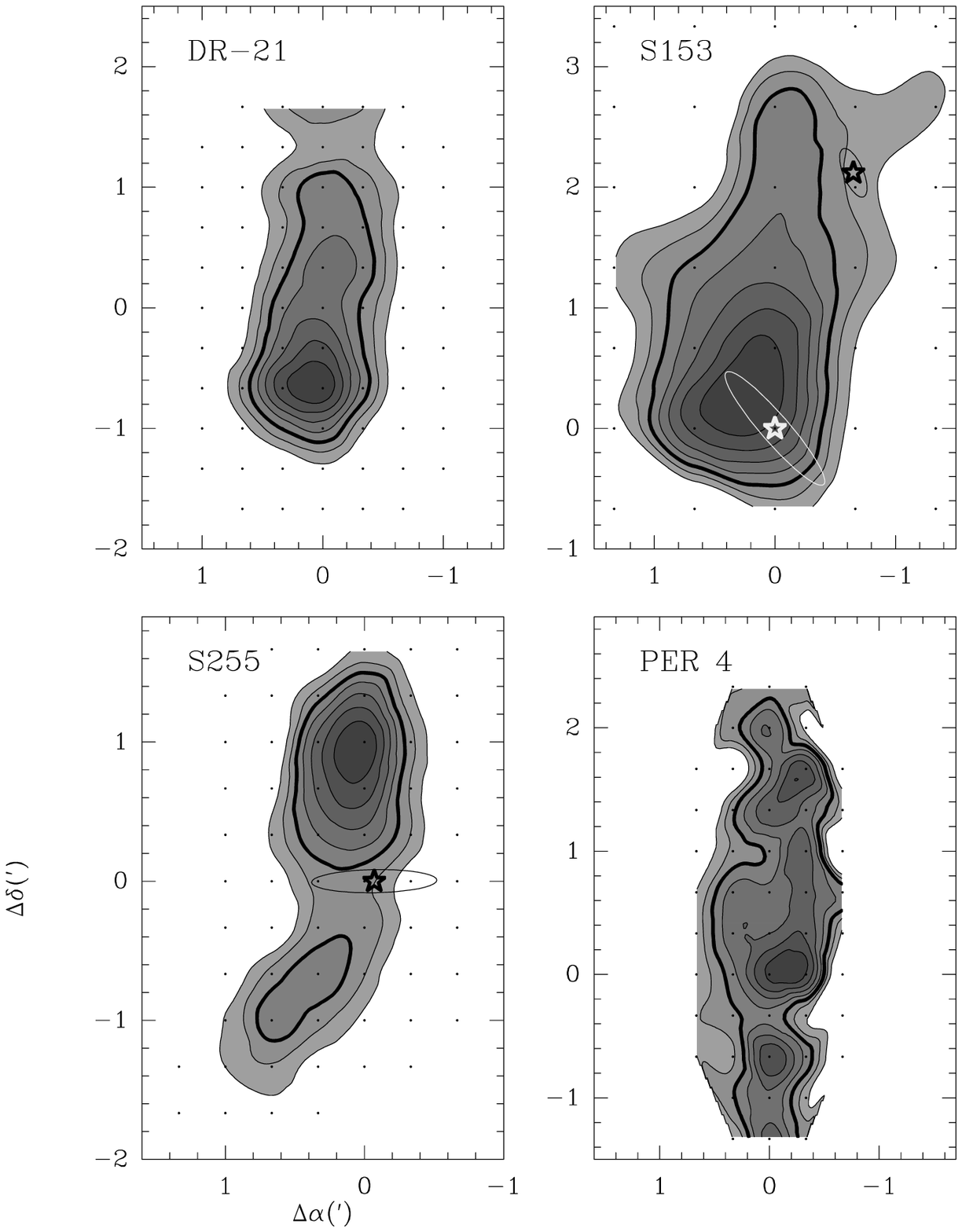}
\caption{Continued}
\end{figure}

\addtocounter{figure}{-1}

\begin{figure}
 \centering \includegraphics[width=9cm]{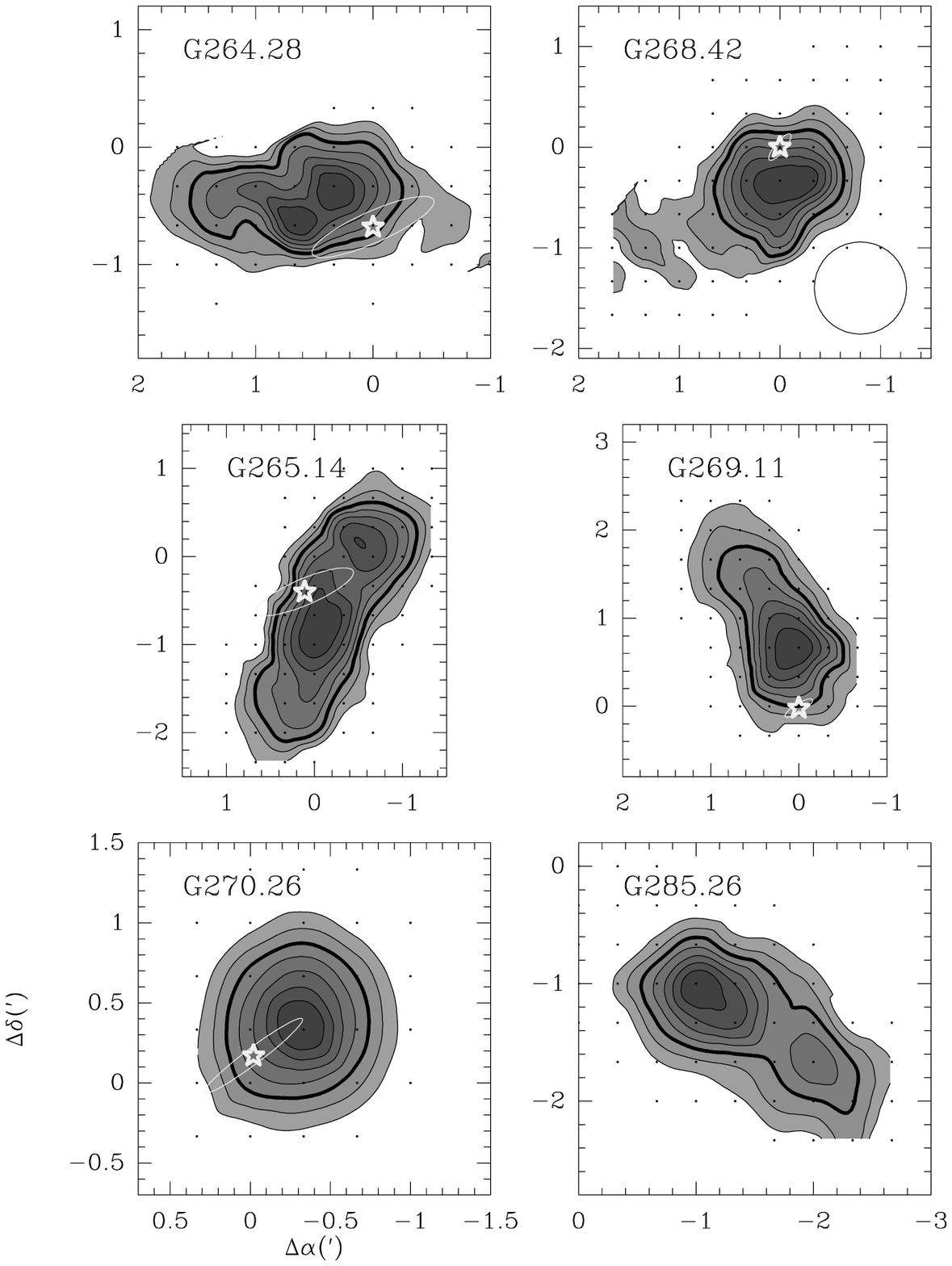}
\caption{Continued}
\end{figure}

\addtocounter{figure}{-1}

\begin{figure}
 \centering \includegraphics[width=9cm]{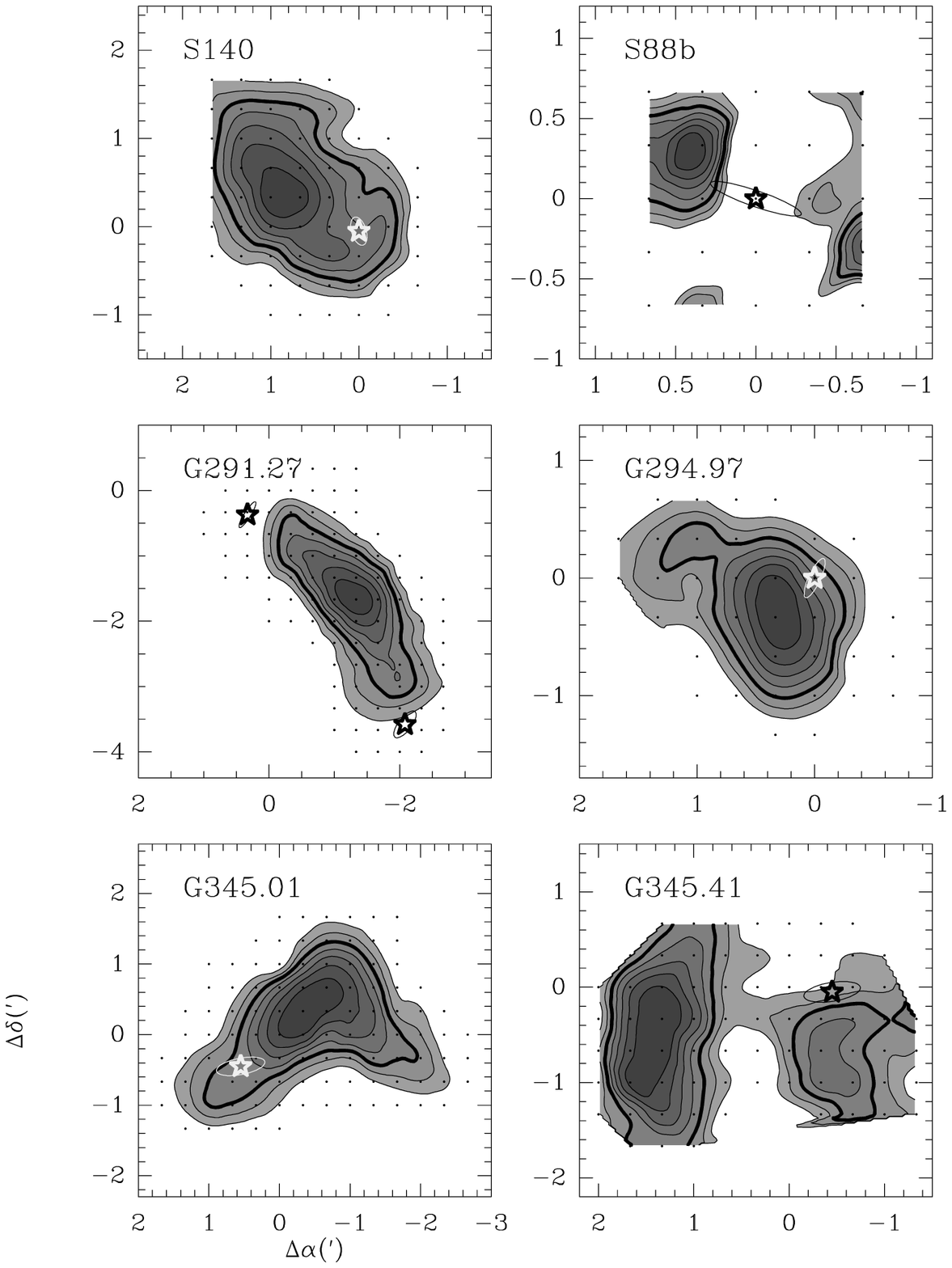}
\caption{Continued}
\end{figure}

\addtocounter{figure}{-1}

\begin{figure}
 \centering \includegraphics[width=9cm]{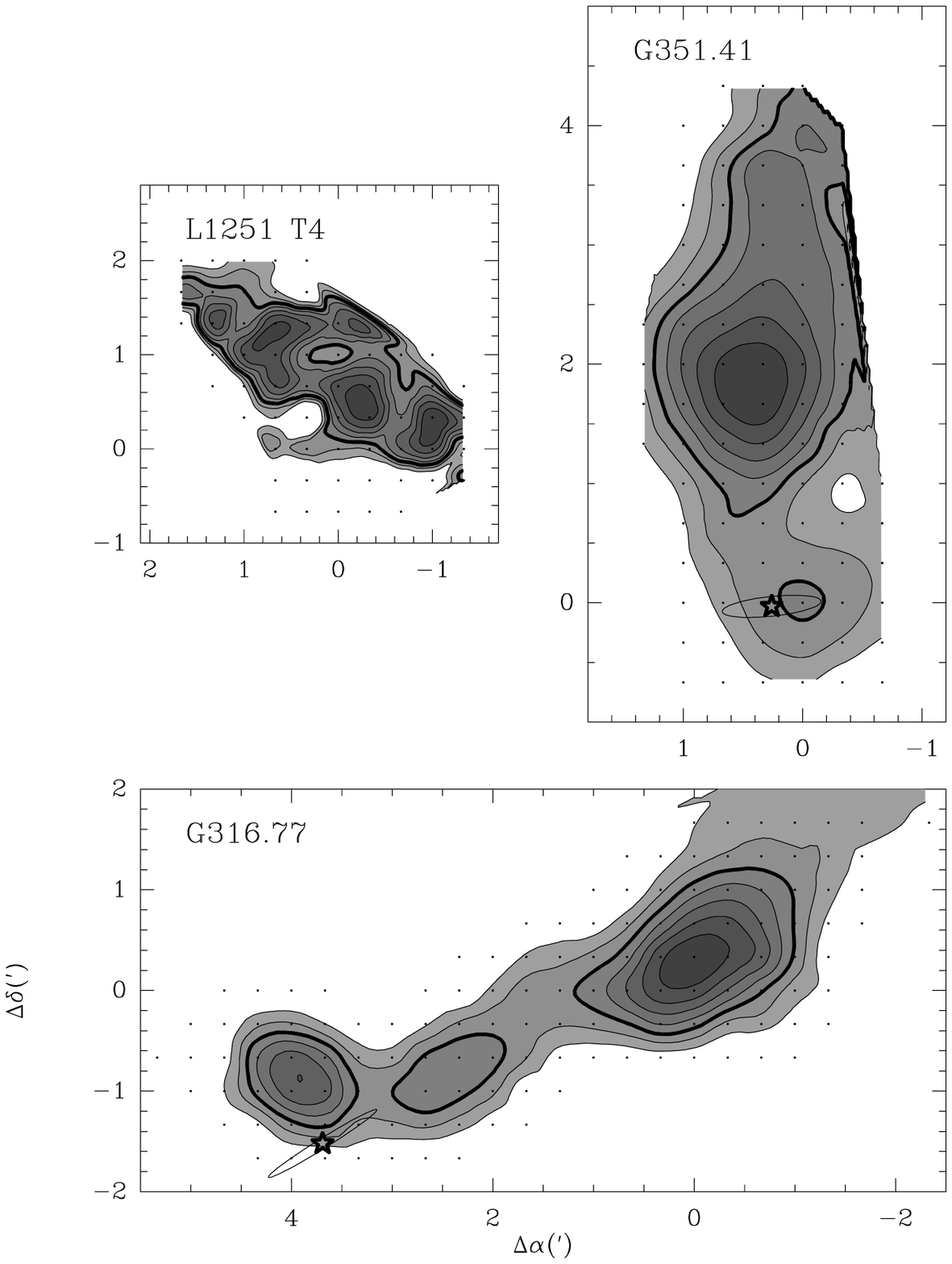}
\caption{Continued}
\end{figure}

Peak integrated intensities are given in Table~\ref{table:line}.
For the sources having separate clumps
at the 30\% level, peak intensity for each clump is given.
The uncertainties of integrated intensities have been calculated as
$\Delta T_{\rm MB} \sqrt{N_{\rm ch}}\,\delta V_{\rm ch}$, where
$\Delta T_{\rm MB}$ -- noise level in channel without line (after baseline
subtracting), $N_{\rm ch}$ -- number of line channels,
$\delta V_{\rm ch}$ -- velocity resolution.
For the sources observed at OSO we used high-resolution autocorrelator data,
with the exception of S~88B for which only filterbank data are available.

\begin{table*}
\centering
\caption[]{N$_2$H$^+$(1--0) line parameters and LTE column densities}
\small
\begin{tabular}{lrrrrrrr}
\noalign{\hrule}\noalign{\smallskip}
Source ($\Delta \alpha \arcsec$, $\Delta \delta \arcsec$)
         &$I$            & $T_{\rm MB}$(23--12) &  $R_{12}$ & $R_{02}$ & $V_{\rm LSR}$(23--12) & $\Delta V$     & $N$(N$_2$H$^+$)  \\
         &(K km s$^{-1}$)&  (K)                 &           &          &  (km s$^{-1}$)        &  (km s$^{-1}$) & (cm$^{-2}$)      \\
         &               &                      &           &          &                       &                & ($10^{12}$) \\
\noalign{\smallskip}\hline\noalign{\smallskip}

RNO~1B (0,0)        & 27.0(0.4)   & 2.88(0.04) & 0.66(0.01) & 0.47(0.02) & --17.55(0.01) & 2.39(0.03) & 31.1\\
NGC~281 (0,0)       & 27.2(0.4)   & 2.29(0.03) & 0.68(0.02) & 0.48(0.02) & --30.36(0.02) & 2.91(0.04) & 31.2\\
S~187 (0,0)         &  6.3(0.4)   & 1.6(0.1)   & 0.68(0.07) & 0.57(0.08) & --13.33(0.03) & 0.91(0.06) &  7.2\\
S~187 (120,--80)    &  4.9(0.4)   & 0.73(0.07) & 0.7(0.1)   & 0.6(0.1)   & --13.59(0.05) & 1.5(0.2)   &  5.6\\
G~133.69+1.22 (160,--160)& 11.7(0.7)   & 1.4(0.1)   & 0.62(0.07) & 0.6(0.1)   & --38.38(0.05) & 1.8(0.1) & 13.4\\
G~133.69+1.22 (40,--80)  &  8.0(0.7)   & 0.38(0.05) & 1.0(0.2)   & 0.8(0.2)   & --41.9(0.2)   & 3.6(0.4) &  9.2\\
S~199 (0,--20)      &  9.7(0.6)   & 0.94(0.06) & 0.64(0.07) & 0.6(0.1)   & --38.02(0.06) & 2.3(0.2)   & 11.2\\
S~201 (40,40)       &  3.9(0.4)   & 0.26(0.03) & 1.2(0.2)   & 0.8(0.2)   & --37.3(0.1)   & 2.8(0.3)   &  4.5\\
AFGL~490 (--20,0)   &  7.6(0.6)   & 1.5(0.1)   & 0.67(0.09) & 0.6(0.1)   & --12.24(0.04) & 1.1(0.1)   &  8.7\\
G142.00 (--20,--80) & 10.9(0.7)   & 1.2(0.1)   & 0.7(0.1)   & 0.4(0.1)   & --13.00(0.06) & 1.8(0.2)   & 12.5\\
Per~4  (0,0)        &  4.5(0.3)   & 2.07(0.09) & 0.69(0.05) & 0.58(0.05) &    7.62(0.01) & 0.48(0.02) &  5.2\\
G~173.48+2.45 (0,0)      & 43.4(0.5)   & 3.78(0.04) & 0.72(0.01) & 0.53(0.02) & --16.12(0.01) & 2.70(0.03) & 49.9\\
G~173.58+2.44 (0,--20)   & 12.2(0.4)   & 1.71(0.06) & 0.74(0.04) & 0.56(0.05) & --16.42(0.02) & 1.63(0.07) & 14.0\\
AFGL~6366 (0,0)     & 14.7(0.3)   & 1.56(0.03) & 0.70(0.02) & 0.56(0.03) &    2.68(0.02) & 2.16(0.05) & 16.9\\
AFGL~6366 (80,60)   & 10.5(0.4)   & 1.45(0.06) & 0.67(0.05) & 0.53(0.06) &    2.10(0.03) & 1.63(0.08) & 12.0\\
S~255 (0,60)        & 27.5(0.6)   & 2.58(0.05) & 0.67(0.02) & 0.53(0.03) &    8.97(0.02) & 2.61(0.06) & 31.6\\
S~76E (--20,--20)   & 79.3(0.7)   & 6.03(0.05) & 0.70(0.01) & 0.53(0.02) &   32.44(0.01) & 3.10(0.03) & 91.1\\
S~88B (20,20)       &  4.3(0.8)   & 0.5(0.2)   & 0.6(0.4)   & 0.7(0.4)   &   22.7(0.1)   & 1.3(0.4)   &  5.6\\
DR~21 (0,--40)      &106.3(0.7)   & 6.93(0.04) & 0.72(0.01) & 0.66(0.01) &  --3.25(0.01) & 3.51(0.02) & 122.0\\
S~140 (60,20)       & 48.2(0.7)   & 6.53(0.09) & 0.64(0.02) & 0.56(0.02) &  --7.03(0.01) & 1.73(0.03) &  55.4\\
L~1251T4 (--60,20)    &  3.1(0.3)   & 1.0(0.1)   & 0.65(0.13) & 0.37(0.13) &  --3.81(0.03) & 0.49(0.05) & 3.6\\
S~153 (0,40)        & 21.2(1.0)   & 2.0(0.1)   & 0.76(0.06) & 0.55(0.08) & --51.23(0.05) & 2.35(0.12) &  24.4\\
G~264.28$+$1.48 (20,--20) & 6.2(0.1) & 0.64(0.01) & 0.70(0.02) & 0.51(0.02) &   6.64(0.01) & 2.17(0.04)  &  7.1\\
G~265.14$+$1.45 (0,--60)  & 15.8(0.2) & 1.61(0.02) & 0.75(0.01) & 0.57(0.02) &   7.40(0.01) & 2.11(0.03) & 18.2\\
G~267.94$-$1.06 (0,20)    &  3.4(0.2) & 0.56(0.04) & 0.53(0.07) & 0.37(0.08) &   3.29(0.04) & 1.36(0.12) &  3.9\\
G~268.42$-$0.85 (0,--20)  & 10.9(0.2) & 1.13(0.02) & 0.66(0.02) & 0.44(0.03) &   3.40(0.02) & 2.56(0.06) & 12.5\\
G~269.11$-$1.12 (0,40)    & 24.3(0.2) & 1.66(0.01) & 0.76(0.01) & 0.56(0.02) &  10.49(0.01) & 3.23(0.03) & 28.0\\
G~270.26$+$0.83 (--20,20) & 20.9(0.2) & 1.44(0.02) & 0.73(0.01) & 0.49(0.02) &   9.69(0.02) & 3.42(0.03) & 24.0\\
G~285.26$-$0.05 (--60,--60) & 12.4(0.2) & 0.92(0.01) & 0.63(0.02) & 0.55(0.02) &   2.54(0.02) & 3.12(0.04) & 14.1\\
G 291.27$-$0.71 (--80,--100)& 36.1(0.2) & 2.91(0.01) & 0.69(0.01) & 0.53(0.01) &--24.36(0.01) & 3.00(0.02) & 41.6\\
G~294.97$-$1.73 (20,--20)   & 15.6(0.2) & 1.53(0.02) & 0.69(0.01) & 0.41(0.02) & --8.05(0.01) & 2.53(0.03) & 17.9\\
G~305.36$+$0.15 (20,40)     &  7.0(0.2) & 0.47(0.02) & 0.69(0.05) & 0.64(0.07) &--38.54(0.05) & 3.2(0.1)   &  8.1\\
G~316.77$-$0.02 (0,20)    & 51.6(0.2) & 3.53(0.01) & 0.72(0.01) & 0.60(0.01) &--39.56(0.01) & 3.39(0.01)   & 59.3\\
G~345.01$+$1.80 (--20,20)  & 58.5(0.2) & 4.07(0.02) & 0.65(0.01) & 0.56(0.01) &--14.08(0.01) & 3.55(0.01)  & 67.3\\
G~345.41$-$0.94 (80,--20)  & 13.7(0.2) & 1.05(0.01) & 0.64(0.01) & 0.47(0.02) &--22.31(0.02) & 3.25(0.04)  & 15.7\\
G~351.41$+$0.64 (20,100)   & 128.7(0.2) & 6.35(0.03) & 0.78(0.01) & 0.79(0.01) & --4.23(0.01) & 4.38(0.02) &148.0\\
\noalign{\smallskip}\hline\noalign{\smallskip}
\end{tabular}
\label{table:line}
\end{table*}

\normalsize

\subsection{N$_2$H$^+$(1--0) line parameters and peak column densities}
\label{sec:line}

\begin{figure}
\centering
\includegraphics[width=9cm]{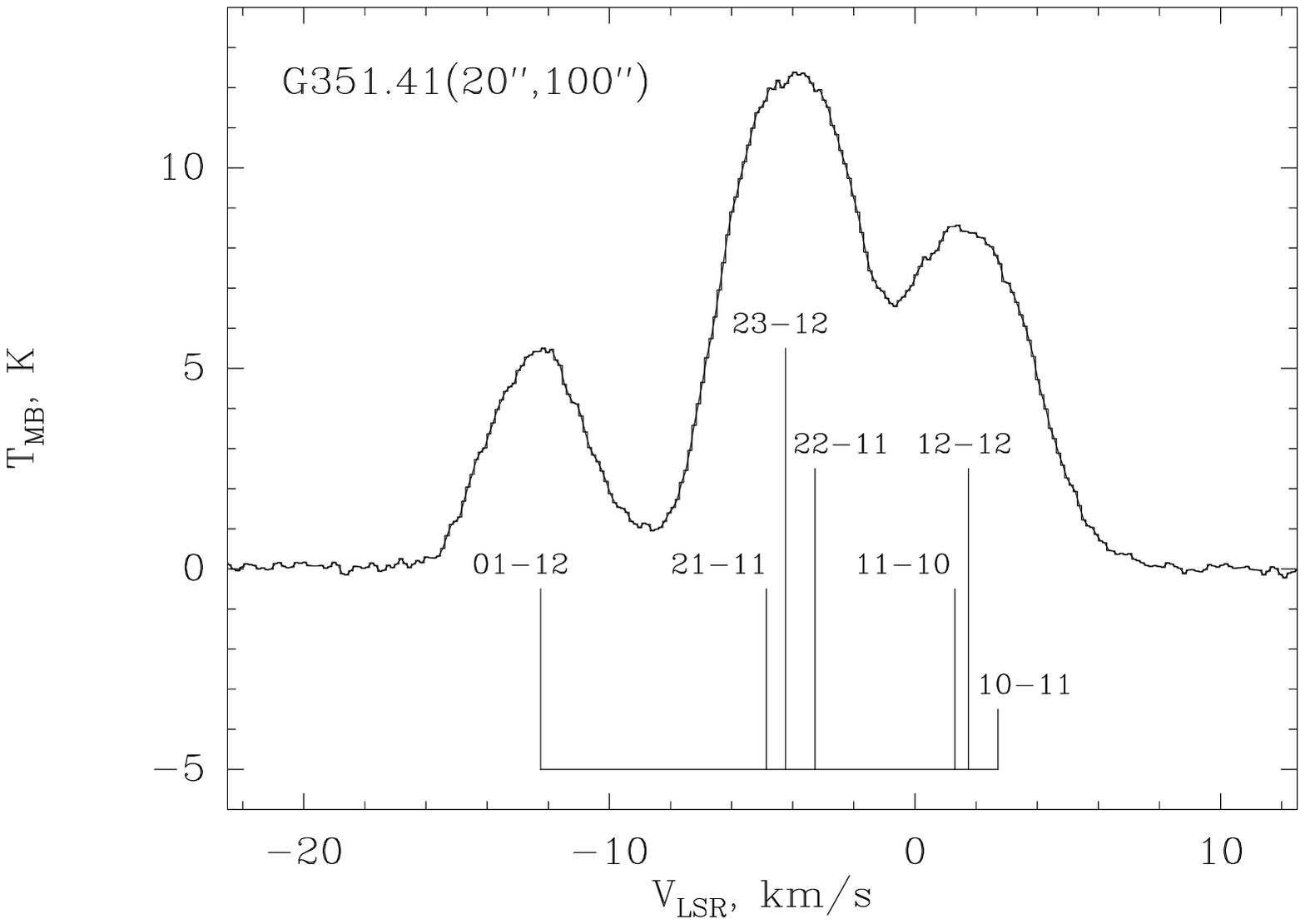}
\caption{The N$_2$H$^+$(1--0) spectrum toward the intensity peak
of G~351.41$+$0.64. The positions of hyperfine components
are indicated. The relative amplitudes of the components correspond
to the LTE optically thin case}
\label{g351}
\end{figure}

N$_2$H$^+$(1--0) spectra consist of seven hyperfine components
(Womack et al. \cite{womack}, Caselli et al. \cite{caselli1}).
However, due to overlap of closely spaced components
only three lines (two of which are actually triplets)
have been observed in most cases (with exception of L1251 and Per4).
As an example, the N$_2$H$^+$(1--0) spectrum
toward the peak position in G~351.41$+$0.64
is shown in Fig.~\ref{g351} with hyperfine components
indicated.
The relative amplitudes of the components correspond to the LTE
optically thin case.
Each component is marked by two pairs of numbers $(F_1^{\prime}~F^{\prime}-F_1~F)$
corresponding to initial and final states of the transition
(Caselli et al. \cite{caselli1}).

After removing baselines from the data (low-order polynomials)
we performed gaussian fitting.
While choosing three gaussians for fitting the data can give only crude
information about position and amplitudes of the components,
fitting by seven independent gaussians with fixed separations
usually leads to incorrect results
(negative amplitudes for some of the overlapped components).
We decided to fit the data by seven gaussians,
only three of which having
independent amplitudes, whereas the amplitudes of the overlapping
components have been assumed proportional to their statistical weights.
This method proved to be useful for treating both LTE and non-LTE cases
and gave lowest $\chi^2$ values in most cases.

\begin{figure}
\centering
\includegraphics[width=9cm]{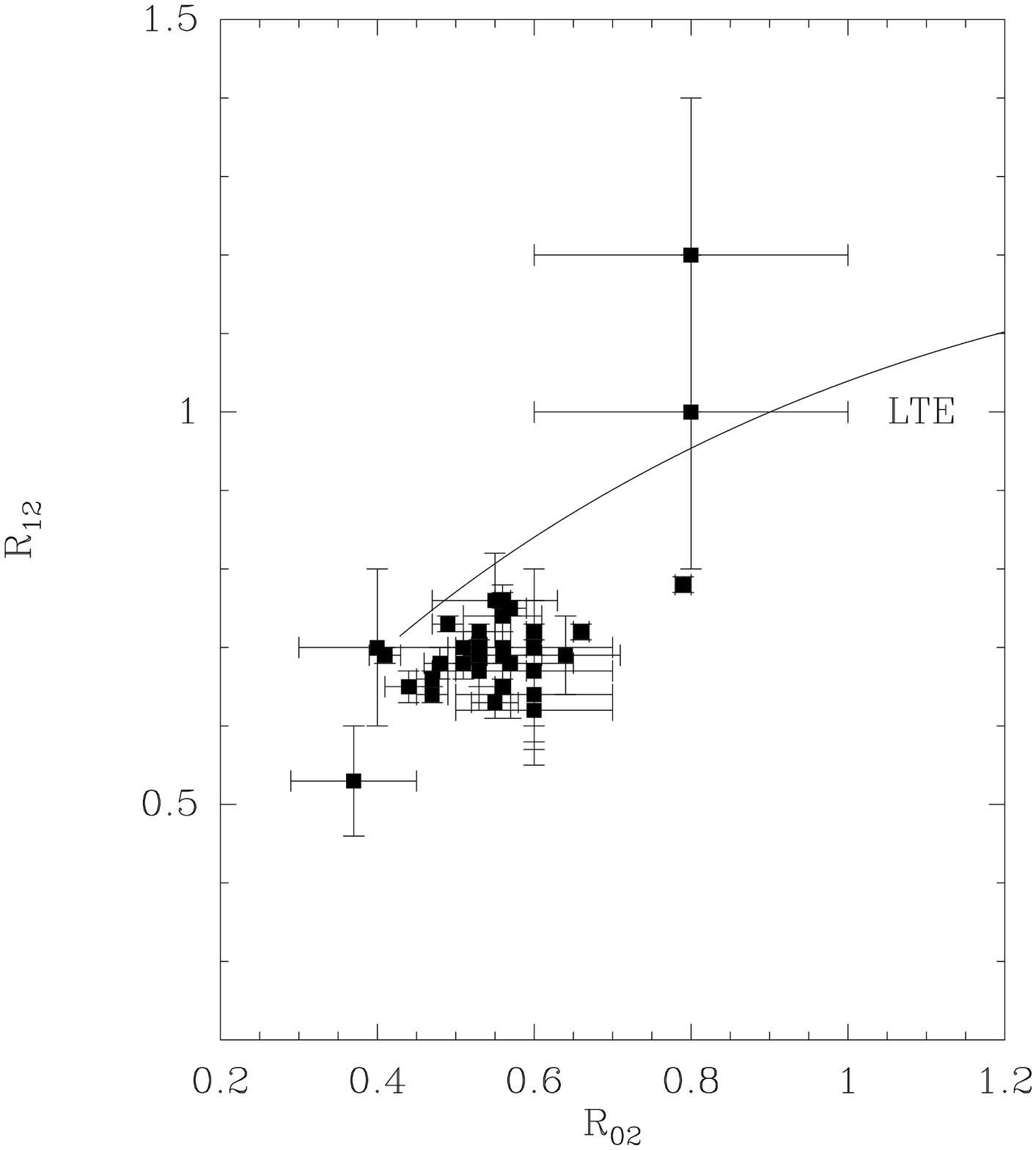}
\caption{Ratio of (12--12) to (23--12) peak temperatures versus ratio of
(01--12) to (23--12) peak temperatures ($R_{12}$ and $R_{02}$, respectively).
The LTE curve is also shown}
\label{ratios}
\end{figure}

The derived parameters are (Table~\ref{table:line}): main beam brightness
temperature of the (23--12) component,
peak intensity ratios of the (12--12) to (23--12) and (01--12) to (23--12)
transitions ($R_{12}$ and $R_{02}$, respectively), velocity
of the (23--12) component and line width at half maximum level,
considered equal for all components.
For most of the sources the $R_{12}$ ratios do not significantly differ from
the optically thin value (5:7) while the $R_{02}$ ratios are usually higher
than expected in the optically thin case (3:7).
Observed $R_{12}$ versus $R_{02}$ ratios together with the LTE curve
are shown in Fig.~\ref{ratios}.
The mean ratios for the sources given in Table~\ref{table:line}
are: $\langle R_{12}\rangle =0.70(0.01)$ and
$\langle R_{02}\rangle =0.57(0.03)$.
Similar ratios apply for each source when all map positions with significant
emission are averaged, i.e., $\langle R_{12}\rangle$ is close to 5:7
while $\langle R_{02}\rangle$ is higher than 3:7.
Fitting by three independent gaussians also reveals an enhanced
intensity of the (01--12) component compared with the LTE case.

No evidence of broad line wings as indicators of outflow
activity has been found in the spectra.
The N$_2$H$^+$(1--0) spectra towards several positions of
G~316.77, G~345.01 and G~351.41
are highly non-gaussian apparently associated
with more than one clump along the line of sight.
Fitting these spectra by the procedure described above
resulted in much higher line widths than for the rest of spectra.
No attempts have been made to decompose these spectra
into several velocity components due to the ambiguity of such a procedure,
and we have excluded these sources from our velocity analysis.

In order to derive line optical depths in the case
of closely spaced and overlapping components
the following function could be used:

\begin{equation}
T(\nu)=\bigl [ \sum_i J(T_{\rm EX_i})\frac{\tau_i(\nu)}{\sum_k \tau_k(\nu)}-
J(T_{BG}) \bigr ] \cdot
\bigl [ 1-\exp(-\sum_k \tau_k(\nu)) \bigr ] ,
\label{eq:fit}
\end{equation}

\noindent{where $J(T)=\frac{h\nu}{k}[\exp(\frac{h\nu}{kT})-1]^{-1}$,
$T_{\rm BG}=2.7~K$ is the cosmic background radiation temperature,
$T_{\rm EX_i}$ and $\tau_i(\nu)$ are excitation temperature and optical depth
of the $i$-th component at the frequency $\nu$, respectively, and summing
is going over all components.}
For the case of the observed N$_2$H$^+$(1--0) spectra
the fitting function (\ref{eq:fit}) could be simplified.
It is known that excitation temperatures of the overlapping lines
always tend to be equal due to photon exchanges
(e.g. Guilloteau et al. \cite{G81}).
Assuming equal excitation temperatures for the overlapping components
within triplets, and, that the (01--12) line
and the triplets do not overlap, the fitted function becomes:

\begin{equation}
T(\nu)=\bigl [ \sum_i J(T_{\rm EX_i})-J(T_{BG}) \bigr ] \cdot
\bigl [1-\exp(-\sum_k \tau_{ik}(\nu)) \bigr ] \hspace{2mm} ,
\label{eq:fit1}
\end{equation}

\noindent{where the first sum is going over non-overlapping groups of components
and the second is going over overlapping components within individual group.}
$\tau_{ik}(\nu)$ is the optical depth
of the $i$-th group and $k$-th overlapping component at the frequency $\nu$.
Thus, the fitted parameters are: three excitation temperatures,
optical depth of the central (23--12) component as well as line width
and velocity of the central component.

The spectra toward the positions given
in Table~\ref{table:line} as well as toward several additional emission peaks
have been fitted by the function (\ref{eq:fit1}).
In three sources with low S/N ratios
(G~133.69, S~88B and L~1251) the solutions did not converge.
In 10 cases the derived optical depth exceeds 3$\sigma$.
For the remaining sources an upper 3$\sigma$ level has been derived
and it exceeds unity in 8 cases (in the case of S~201
$\tau$(23--12)$<$3, the highest upper limit among the sources).
The results of fitting for 10 sources (excitation temperatures,
velocities, line widths and optical depths)
are given in Table~\ref{table:optdepth}.
The optical depth and excitation temperature ranges for different sources
are: $\sim 0.2-1$ and $\sim 7-29$~K, respectively.
The difference in excitation temperature between different groups
of components is $\le 5.6$~K and in several cases
$T_{\rm EX}$(01--12) is the highest one.
As the sources given in Table~\ref{table:optdepth} are among
those with highest intensities we conclude
that $T_{\rm EX}\la 25$~K and $\tau (23-12)\la 1$ for the studied sample.
It should be emphasized that we have assumed a beam filling factor
of 1 in all cases; if smaller, excitation temperatures are higher.

Since non-LTE effects seem small, we have assumed LTE and the optically thin approximation
to calculate N$_2$H$^+$ column densities, $N$(N$_2$H$^+$),
for all sources in Table~\ref{table:line}.
Excitation temperatures were taken equal to 10~K for all sources,
the value at which LTE column densities are close to minimum.
At $T_{\rm EX}=20$~K column densities are 1.6 times higher.
The calculated values, which could be treated as lower limits,
are given in the last column of Table~\ref{table:line}.

\begin{table*}
\centering
\caption[]{N$_2$H$^+$(1--0) line parameters with optical depths}
\small
\begin{tabular}{lrrrrrr}
\noalign{\hrule}\noalign{\smallskip}
Source ($\Delta \alpha \arcsec$, $\Delta \delta \arcsec$)
       & $T_{\rm EX}$(12--12) & $T_{\rm EX}$(23--12) & $T_{\rm EX}$(01--12)  & $V_{\rm LSR}$(23--12) & $\Delta V$   & $\tau$(23--12)  \\
       &      (K)             &      (K)             &       (K)             &  (km s$^{-1}$)        & (km s$^{-1}$) \\
       &                      &                      &                       &                       &               \\
\noalign{\smallskip}\hline\noalign{\smallskip}

RNO~1B (0,0)        & 7.75(0.37) & 9.18(0.38) & 7.32(0.50) & --17.54(0.01) & 1.94(0.06) & 1.0(0.1) \\
S~255 (0,60)        &  9.3(1.6)  & 10.6(1.6)  & 10.1(2.2)  &    8.96(0.02) & 2.31(0.10) & 0.5(0.2) \\
S~76E (--20,--20)   & 24.4(4.2)  & 27.0(4.3)  & 27.1(5.4)  &   32.44(0.01) & 2.86(0.06) & 0.3(0.1) \\
DR~21 (0,--40)      & 20.3(1.3)  & 22.8(1.3)  & 25.0(2.1)  &  --3.25(0.01) & 3.09(0.04) & 0.6(0.1) \\
S~140 (60,20)       & 21.4(3.7)  & 25.4(4.1)  & 27.0(5.5)  &  --7.03(0.01) & 1.62(0.05) & 0.4(0.1) \\
G~270.26$+$0.83 (--20,20) & 8.8(1.5) & 9.2(1.5) & 8.8(1.8) &    9.69(0.02) & 3.17(0.08) & 0.3(0.1) \\
G~291.27$-$0.71 (--80,--100)& 24.2(6.0) & 25.9(6.2) & 29.1(7.7) & --24.36(0.01) & 2.90(0.03) & 0.15(0.05) \\
G~316.77$-$0.02 (240,--60)  &  8.9(0.6) & 10.1(0.6) & 10.1(0.9) & --37.94(0.01) & 3.61(0.06) & 0.5(0.1) \\
G~345.01$+$1.80 (--20,20) & 11.8(0.3) & 14.3(0.3) & 13.4(0.5)   & --14.07(0.01) & 3.08(0.02) & 0.62(0.03) \\
G~351.41$+$0.64 (20,100)  & 16.3(0.3) & 18.2(0.2) & 20.6(0.5)   & --4.20(0.01)  & 3.65(0.02) & 0.86(0.03) \\
\noalign{\smallskip}\hline\noalign{\smallskip}
\end{tabular}
\label{table:optdepth}
\end{table*}

\normalsize

\subsection{Clump deconvolution}
\label{sec:deconv}

Most of the maps in Fig.~\ref{maps} show elongated or more complex
structures with prominent secondary peaks.
In order to estimate core sizes, two-dimensional (2D)
gaussian fittings have been used.
As a first step we tried to fit simultaneously a number of circular
2D gaussians to each map in order to reveal distinct clumps.
The initial guesses of number of components and their locations
were simply based on inspections of the maps in Fig.~\ref{maps} and
strip scans, obtained where required.
In the second step the circular gaussians were replaced by elliptical
ones, adding two more parameters to be determined: axial ratio and
position angle. Finally, core sizes were estimated by deconvolving the
geometric mean of the extents of the fitted elliptical gaussians with
the beam size, i.e., $C=\sqrt{G^2-B^2}$ where $C$, $G$, and $B$ are the full
half widths of the core, observed geometric mean of the extents and beam, respectively.
In general, 1--2 clumps per map were found but in several sources (S~187,
G~291.27, G~316.77, G~345.01 and G~351.41)
three clumps have been revealed.
We have excluded Per~4 and L~1251T4 from this analysis,
(the former has already been analyzed in Paper~I)
as well as sources with small ($\le 25$ points per map)
and incomplete maps.
In total, 47 clumps in 26 objects have been revealed.
The relative coordinates of the clumps, their axial ratios,
angular and linear sizes calculated at half maximum level
are given in Table~\ref{table:phys} together with fitting uncertainties.
We have not taken into account possible uncertainties in source distances
while calculating linear sizes.
Clumps in the same core are marked by numbers.
Angular sizes of some clumps are smaller than the telescope beam.
These clumps and those having high uncertainties in size
have been excluded from calculations of physical parameters
(Section~\ref{sec:phys}).
For four cores where all revealed clumps are smaller than the telescope beam
the parameters of the whole core are also given in Table~\ref{table:phys}.
For three of them (G~265.14, G~269.11 and G~285.26) physical parameters
have been calculated.

\subsection{Physical parameters of clumps}
\label{sec:phys}

In column 7 of Table~\ref{table:phys} mean line widths
averaged over the half maximum intensity regions are given.
For overlapping clumps that are not separated at this level,
line widths have been averaged over the whole regions.
The following formula was used for weighted averages:

\begin{equation}
\langle\Delta V\rangle=\frac{\sum w_i\Delta V_i}{\sum w_i} \hspace{2mm},
\label{eq:vmean}
\end{equation}

\noindent{where $\Delta V_i$ -- velocity dispersion in the $i$-th position,
$w_i=\sigma_i^{-2}$ and $\sigma_i$ -- uncertainty of velocity
dispersion in the $i$-th position calculated from the fit.}
Uncertainties of mean line width were calculated as:

\begin{equation}
\sigma_{\langle\Delta V\rangle}=\sigma
\frac{\sqrt{\sum w_i^2}}{\sum w_i} \hspace{2mm},
\end{equation}

\noindent{where}

\begin{equation}
\sigma=\sqrt{\frac{\sum w_i\Delta V_i^2}{\sum w_i}
- \langle\Delta V\rangle ^2}
\label{eq:sigmav}
\end{equation}

\noindent{is the uncertainty of individual line width measurements taking
into account point-to-point differences in line widths and using
the same weights.}

Mean line widths vary from 1~km~s$^{-1}$ to 4.7~km~s$^{-1}$, i.e. much higher
than thermal widths. So these values should be considered
as non-thermal velocity dispersion in the sources without any recalculations.

Following Paper~I we have calculated virial masses
of homogeneous spherically-symmetrical clumps with no external pressure
and no magnetic field (e.g. Pirogov \& Zinchenko \cite{pz}):

\begin{equation}
M_{\rm vir}(M_{\odot})=105 \langle\Delta V\rangle^2 \cdot d \hspace{2mm},
\label{eq:mvir}
\end{equation}

\noindent{where $\Delta V$ and $d$ are in km~s$^{-1}$ and pc units.}
Virial masses (column 8 of Table~\ref{table:phys}) lie in the range:
$\sim 30-3000$~$M_{\odot}$.
These values should be treated as upper limits.
If clumps have inner density profiles $\sim r^{-\alpha}$,
equation (\ref{eq:mvir}) should be multiplied
by the factor $\frac{3}{5}\cdot\frac{5-2\alpha}{3-\alpha}$
which for $\alpha\ge 0$ is $\le 1$.
Mean volume densities for spherically-symmetric clumps have been calculated
as

\begin{equation}
n_{\rm vir}=3 M_{\rm vir}/(4\pi r^3 m) \hspace{2mm},
\end{equation}

\noindent{where $r=d/2$ is the radius of the half maximum intensity region,
$m=2.33$~amu, mean molecular mass.}
Mean volume densities lie in the range: $(0.5-17.1)\cdot 10^4$~cm$^{-3}$
(see column 9 of Table~\ref{table:phys}).
The N$_2$H$^+$ abundances are given in column 10 of Table~\ref{table:phys}:

\begin{equation}
X({\rm N}_2{\rm H}^+)=\frac{\langle N({\rm N}_2{\rm H}^+)\rangle}
{\langle N({\rm H}_2)\rangle} \hspace{2mm},
\end{equation}

\noindent{where $\langle N($N$_2$H$^+)\rangle$ is mean column density
averaged over the emission region defined by $I/I_{\rm MAX}\ge 0.5$ and
$\langle N({\rm H}_2)\rangle$ is molecular hydrogen column density
calculated for the same region using virial mass estimate.}
Calculated abundances lie in the range: $(1.2-12.8)\cdot 10^{-10}$.
Averaged over 36 analyzed clumps,
$\langle X($N$_2$H$^+)\rangle=(5.2\pm 0.5)\cdot 10^{-10}$.

\begin{table*}
\centering
\caption[]{Physical parameters of N$_2$H$^+$ clumps}
\small
\begin{tabular}{lrrrrrrrrr}
\noalign{\hrule}\noalign{\smallskip}
Source   & $\Delta\alpha$  & $\Delta\delta$ & Axial  &$\Delta \Theta$ & $d$ & $\langle\Delta V\rangle$ &$M_{\rm vir}$   & $n_{\rm vir}$    & $X$(N$_2$H$^+$)     \\
         &  ($\arcsec$)    &  ($\arcsec$)   & ratio  &  ($\arcsec$)   &(pc) & (km s$^{-1}$)            & ($M_{\odot}$)  &  (cm$^{-3}$)     & ($10^{-10}$)   \\
         &                 &                &        &                &     &                          &                & ($10^4$)    \\
\noalign{\smallskip}\hline\noalign{\smallskip}

RNO~1B             &   6(2)  & --1(2)     &1.6(0.1)  & 87(4)    & 0.36(0.02) & 2.0(0.1) &  153& 11.0 & 3.5 \\
NGC~281            &   7(1)  & --7(1)     &1.6(0.2)  & 42(3)    & 0.72(0.05) & 2.9(0.1) &  611&  5.6 & 4.7 \\
S~187 (1)          &   6(7)  &   2(7)     &1.7(0.7)  & 63(13)   & 0.31(0.07) & 1.0(0.1) &   29&  3.5 & 3.5 \\
S~187 (2)          & 199(11) & --79(10)   &6.3(29.9) & 29(69)  \\
S~187 (3)          & 115(8)  & --98(9)    &1.4(0.6)  & 81(17)   & 0.39(0.08) & 1.6(0.3) &  107&  5.9 & 1.2 \\
G~133.69+1.22 (1)  &  31(4)  & --54(5)    &1.9(0.5)  & 75(9)    & 0.76(0.09) & 3.2(1.3) &  793&  6.0 & 1.2 \\
G~133.69+1.22 (2)  & 171(5)  & --168(6)   &2.3(1.0)  & 52(11)   & 0.53(0.12) & 2.0(0.4) &  213&  4.7 & 2.8 \\
S~199              &  12(4)  & --14(4)    &1.1(0.2)  & 82(9)    & 0.83(0.09) & 1.4(0.1) &  167&  1.0 & 6.0 \\
S~201              &  31(2)  &   19(3)    &1.7(0.5)  & 45(7)    & 0.45(0.07) & 1.6(0.2) &  123&  4.4 & 1.4 \\
G~173.48+2.45      & --3(1)  & --9(1)     &1.7(0.2)  & 54(3)    & 0.60(0.03) & 2.6(0.1) &  411 & 6.3 & 6.7 \\
G~173.58+2.44      & --6(3)  & --12(2)    &1.6(0.2)  & 72(5)    & 0.80(0.06) & 1.4(0.1) &  171&  1.1 & 7.2 \\
AFGL~6366 (1)      &   2(1)  &   6(1)     &1.8(0.3)  & 38(3)    & 0.37(0.03) & 2.1(0.1) &  170& 11.1 & 2.6 \\
AFGL~6366 (2)      &  89(2)  &   57(2)    &1.5(0.3)  & 48(5)    & 0.47(0.05) & 1.5(0.1) &  105&  3.5 & 3.9 \\
S~255 (1)          &   6(1)  &   46(2)    &2.7(0.3)  & 49(3)    & 0.60(0.04) & 2.6(0.1) &  433&  6.8 & 4.5 \\
S~255 (2)          &  23(2)  & --45(3)    &3.0(0.5)  & 56(5)    & 0.68(0.06) & 1.7(0.1) &  212&  2.3 & 6.3 \\
S~76E              & --27(1) & --18(1)    &1.3(0.1)  & 65(2)    & 0.67(0.02) & 2.7(0.1) &  518&  5.8 & 11.4 \\
DR~21 (1)          &   6(1)  & --34(1)    &1.4(0.2)  & 49(3)    & 0.72(0.04) & 3.3(0.1) &  826&  7.5 & 12.8 \\
DR~21 (2)          & --4(2)  &   56(3)    &1.7(0.2)  & 75(4)    & 1.09(0.06) & 3.3(0.1) & 1200&  3.1 & 10.2 \\
S~140 (1)          &  61(2)  &   42(2)    &1.2(0.1)  & 89(3)    & 0.39(0.01) & 1.9(0.1) &  140&  7.9 & 7.0 \\
S~140 (2)          &   7(2)  &  --8(2)    &1.3(0.2)  & 65(4)    & 0.28(0.02) & 1.9(0.1) &  108& 15.7 & 3.5 \\
S~153 (1)          &  16(3)  &   19(4)    &1.3(0.2)  & 78(7)    & 1.51(0.13) & 2.5(0.2) &  990&  1.0 & 8.8 \\
S~153 (2)          & --14(7) &  122(6)    &1.5(0.3)  &110(11)   & 2.13(0.22) & 2.5(0.2) & 1400&  0.5 & 5.9 \\
G~264.28$+$1.48 (1)&  16(2)  & --26(2)    &3.4(3.7)  & 29(16) \\
G~264.28$+$1.48 (2)&  80(5)  & --39(7)    &2.8(4.4)  & 36(28) \\
G~264.28$+$1.48    &  33(3)  & --27(1)    &8.6(11.7) & 37(25) \\
G~265.14$+$1.45 (1)&   7(2)  & --78(2)    &3.6(1.5)  & 42(9) \\
G~265.14$+$1.45 (2)& --32(2) &    2(2)    &2.6(0.9)  & 42(8) \\
G~265.14$+$1.45    & --11(2) & --39(3)    &5.1(0.8)  & 73(6)    & 0.60(0.05) & 2.2(0.1) &  307&  4.7& 3.4 \\
G~268.42$-$0.85    &    4(3) & --25(3)    &1.2(0.3)  & 50(7)    & 0.37(0.05) & 1.9(0.1) &  138&  9.0& 2.3 \\
G~269.11$-$1.12 (1) &   4(1) &   34(1)    &1.3(0.2)  & 46(3) \\
G~269.11$-$1.12 (2) &  41(2) &  102(2)    &1.5(0.4)  & 46(6) \\
G~269.11$-$1.12     &  14(2) &   52(2)    &2.9(0.4)  & 75(5)    & 0.95(0.07) & 2.7(0.1) &  717&  2.8& 5.4 \\
G~270.26$+$0.83     & --18(1) &   22(1)   &2.4(0.8)  & 22(4) \\
G~285.26$-$0.05 (1) & --64(1) & --65(1)   &2.2(0.9)  & 27(6) \\
G~285.26$-$0.05 (2) & --124(2)& --107(2)  &2.0(0.8)  & 38(8) \\
G~285.26$-$0.05     & --85(2) & --79(2)   &3.6(0.8)  & 62(7)     & 1.42(0.15) & 2.9(0.1) & 1220& 1.4& 3.5\\
G 291.27$-$0.71 (1) & --110(1)& --170(1)  & 1.2(0.1)   & 72(3)  & 0.94(0.03) & 2.8(0.1) &  777& 3.2& 4.1\\
G 291.27$-$0.71 (2) & --81(1) & --97(1)   & 1.5(0.2)   & 32(2) \\
G 291.27$-$0.71 (3) & --31(1) & --50(3)   & 1.3(0.2)   & 44(3) \\
G~294.97$-$1.73     &  26(3)  & --14(2)   & 2.4(0.6)   & 58(7)  & 0.34(0.04) & 2.4(0.1) &  200& 17.1& 1.7\\
G~316.77$-$0.02 (1) & --6(3)  &   27(2)   & 2.4(0.2)   &108(5)  & 1.63(0.08) & 3.3(0.1) & 1890&  1.5& 11.0\\
G~316.77$-$0.02 (2) &  130(5) & --37(3)   & 6.0(1.2)   & 88(9)  & 1.32(0.14) & 4.7(0.1) & 3030&  4.4& 3.0\\
G~316.77$-$0.02 (3) &  234(1) & --54(1)   & 2.2(0.3)   & 45(3) \\
G~345.01$+$1.80 (1) &   39(2) & --34(2)   & 1.8(0.2)   & 67(4)  & 0.69(0.04) & 3.3(0.1) &  765&  7.9& 3.9\\
G~345.01$+$1.80 (2) & --39(1) &   30(1)   & 1.3(0.1)   & 72(2)  & 0.73(0.02) & 3.3(0.1) &  818&  6.9& 6.9\\
G~345.01$+$1.80 (3) & --116(2)& --23(3)   & 2.6(1.9)   & 34(12) \\
G~345.41$-$0.94 (1) &   80(1) & --31(3)   & 3.4(0.5)   & 81(6)  & 1.10(0.09) & 3.2(0.1) & 1180&  3.0& 2.5\\
G~345.41$-$0.94 (2) & --36(4) & --43(6)   & 1.4(0.3)   & 112(12)& 1.52(0.16) & 2.2(0.1) &  803&  0.8& 4.0\\
G~351.41$+$0.64 (1) &   2(2)  &  223(2)   & 1.4(0.2)   & 76(5)  & 0.63(0.04) & 3.7(0.1) &  884& 12.1& 6.7\\
G~351.41$+$0.64 (2) &  26(1)  &  115(1)   & 1.1(0.1)   & 81(2)  & 0.67(0.02) & 3.7(0.1) &  946& 10.6& 10.0\\
G~351.41$+$0.64 (3) & --1(2)  &    2(2)   & 1.4(0.2)   & 89(5)  & 0.73(0.04) & 3.8(0.2) & 1120&  9.6& 5.3\\
\noalign{\smallskip}\hline\noalign{\smallskip}
\end{tabular}
\label{table:phys}
\end{table*}

\normalsize

\section{Distributions of the observed parameters over the sources}

According to the present theories of star formation radial dependences
of density and velocity dispersion (which is considered to be
related to magnetic field distribution)
are the most important parameters describing the evolution of star forming cores.
In particular, the standard theory of low-mass star formation
from a singular isothermal sphere (Shu \cite{shu}) gives a radial density law
$\propto r^{-2}$ for nearly static envelope while
in the logatropic sphere model for massive star formation
density varies as $r^{-1}$ (McLaughlin \& Pudritz \cite{mclaughlin}).
Velocity dispersion in the latter model is nearly constant in the center
and rises with distance in the outer regions as $\sim r^{1/3}$.

Recent studies of dust continuum emission in star forming low-mass and
massive cores have not revealed significant differences between density
power-law indices ($\alpha$) for these two classes of objects, implying
$\alpha\sim 1.5-2$. Molecular multi-line modeling of massive cores shows somewhat
lower values (see Section~\ref{sec:intro}).
An analysis of the N$_2$H$^+$(1--0) data in low-mass cores
(Paper~I) revealed the following ranges: $\alpha=1.8-2.8$ for cores with
stars, and $\alpha=1.6-1.9$ for cores without stars.

Goodman et al. (\cite{goodman98}) found that velocity dispersion profiles
in low-mass cores are nearly constant within so-called ``coherent regions"
with sizes of $\sim 0.1$~pc, while rising in the outer regions with power-law index
$\ga 0.2$ which is more in line with the logatropic model.
Low-mass cores observed in N$_2$H$^+$(1--0) (Paper~I)
imply ``coherent lengths" of $\sim 0.01$~pc but show
no common trends in the outer regions.
In contrast, molecular line widths in massive cores increase
toward the center (Zinchenko \cite{zin2}, Lapinov et al. \cite{lapinov},
Fontani et al. \cite{fontani})
which could be connected either with optical depth effects or
with an enhancement of dynamical activity of gas in central regions.

In this section we make an attempt to derive N$_2$H$^+$
integrated intensity and line width radial profiles on the cores
which, under certain assumptions, could be connected with density and velocity
dispersion profiles, respectively.
An analysis of velocity gradients in the cores which could give
a clue to their rotational properties is also given.

\subsection{Integrated intensity profiles}
\label{sec:int-prof}

As N$_2$H$^+$ emission is probably optically thin toward most
of the map positions, except peak positions in few of them (see Section~\ref{sec:line}),
their integrated intensity distributions should follow
column density distributions if excitation conditions do not vary
inside the sources.
If in addition N$_2$H$^+$ abundance is constant, the
integrated intensity distribution is directly related to
the hydrogen column density distribution.
Then, for infinite, spherically-symmetric core the power-law index
for volume density profile will be $p+1$, where $p$ is an integrated
intensity power-law index (see e.g. Motte \& Andr\'e \cite{motte}).

\begin{figure}
\centering \includegraphics[width=9cm]{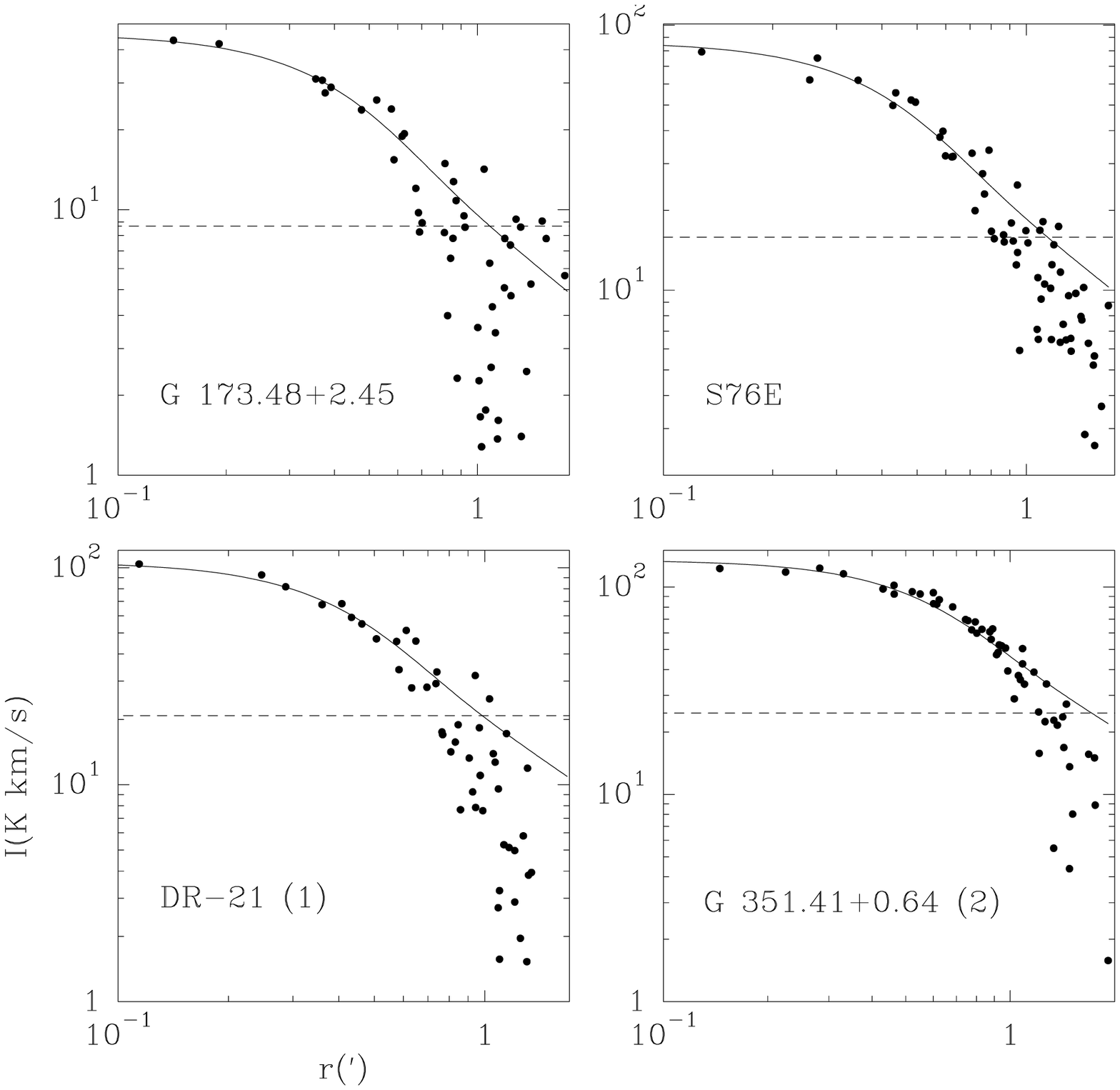}
\caption{Four examples of integrated intensity profiles together with
the fitted power-law functions convolved with a gaussian
telescope beam.
The data lower than the dashed lines (20\% of peak integrated intensity)
were neglected}
\label{int-prof}
\end{figure}

We performed fitting of N$_2$H$^+$(1--0) integrated intensity maps
for 26 clumps which are not highly elongated (axial ratios $<2$)
using  power-law radial distributions
($\propto r^{-p}$) convolved in 2D with a gaussian telescope beam.
Although power laws are not optimal for fitting the
radial distributions (e.g. gaussian functions usually give lower $\chi^2$
values) we used them in order to probe different star formation theories.
The intensity maps integrated over total line range as well as
those integrated over the (01--12) component have been analyzed.
The latter should be less affected by optical depth effects
than total integrated intensity maps, however, due to lower intensities
these maps are more noisy and in several cases fits were not successful.
Power-law index, amplitude and coordinates of maximum position
have been varied in order to get the best fit.
Power-law indices calculated for these two kinds of maps
($p$ and $p'$, respectively) with fitting uncertainties
are given in Table~\ref{table:powerlaw}.
We have also calculated power-law indices for reduced integrated
intensity maps which contain only points with intensities equal or higher
than 20\% and 40\% of peak integrated intensity.
The calculated power-law indices ($p_{20}$ and $p_{40}$, respectively)
and corresponding uncertainties are also given in Table~\ref{table:powerlaw}.
In Fig.~\ref{int-prof} four examples of integrated intensity maps
as 1D profiles are shown together with fitting results
for reduced maps.
The radial distance ($r$) in Fig.~\ref{int-prof} is the distance
from the maximum of fitting curve to the given point of the map.
The data below the dashed lines (20\% of peak integrated intensity)
have not been taken into account during fitting.

\begin{table}[htb]
\centering
\caption[]{ Power-law indices calculated from fits of
 power-law distributions convolved with the telescope beam
to N$_2$H$^+$(1--0) integrated intensity maps with axial ratios $< 2$}
\scriptsize
\begin{tabular}{lrrrr}
\noalign{\hrule}\noalign{\smallskip}
Source   & $p$ & $p'$ & $p_{20}$ & $p_{40}$\\
\noalign{\smallskip}\hline\noalign{\smallskip}

RNO~1B             &  0.91(0.05) & 0.89(0.09) &  0.83(0.05) &   0.68(0.09)\\
NGC~281            &  1.42(0.06) & 1.34(0.16) &  1.17(0.08) &   1.19(0.02)\\
S~187 (1)          &             & 0.83(0.21) &  0.79(0.05) &   0.94(0.09)\\
S~187 (3)          &  1.19(0.22) &            &  0.82(0.14) &   0.76(0.09)\\
G~133.69+1.22 (1)  &  0.91(0.20) &            &  1.37(0.10) \\
S~199              &  0.96(0.11) &            &  0.74(0.11) &   0.39(0.28)\\
S~201              &  1.30(0.15) &            &  1.02(0.20) &   0.67(0.29)\\
G~173.48+2.45      &  1.31(0.04) & 1.40(0.05) &  1.08(0.06) &   0.98(0.08)\\
G~173.58+2.44      &  1.07(0.08) & 0.97(0.26) &  0.84(0.07) &   0.70(0.10)\\
AFGL~6366 (1)      &  1.48(0.03) & 1.20(0.25) &  1.14(0.01) &   1.27(0.26)\\
AFGL~6366 (2)      &  1.43(0.11) & 1.13(0.20) &  1.25(0.08) &   1.21(0.37)\\
S~76E              &  1.22(0.03) & 1.13(0.04) &  1.06(0.05) &   0.93(0.07)\\
DR~21 (1)          &  1.42(0.03) & 1.47(0.03) &  1.12(0.09) &   1.04(0.09)\\
DR~21 (2)          &  1.15(0.08) & 1.14(0.09) &  0.78(0.12) &   0.51(0.20)\\
S~140 (1)          &  0.99(0.05) & 1.02(0.08) &  0.81(0.05) &   0.65(0.08)\\
S~140 (2)          &  1.19(0.06) & 1.11(0.10) &  0.80(0.13) &   0.54(0.14)\\
S~153 (1)          &  1.29(0.14) & 1.30(0.32) &  1.12(0.09) \\
S~153 (2)          &  1.08(0.11) & 1.23(0.25) &  0.66(0.08) &   0.46(0.08)\\
G~268.42$-$0.85    &  1.35(0.10) & 1.58(0.26) &  0.97(0.13) &   0.72(0.31)\\
G~291.27$-$0.71 (1)&  1.33(0.06) & 1.41(0.10) &  1.05(0.05) &   0.94(0.08)\\
G~345.01$+$1.80 (1) & 1.35(0.03) &            &  0.99(0.10) &   0.69(0.15)\\
G~345.01$+$1.80 (2) & 1.37(0.03) & 1.40(0.04) &  1.05(0.06) &   0.76(0.09)\\
G~345.41$-$0.94 (2) & 0.75(0.09) & 0.89(0.17) &  0.72(0.08) &   0.63(0.05)\\
G~351.41$+$0.64 (1) & 1.40(0.08) & 1.41(0.03) &  1.01(0.01) &   0.77(0.04)\\
G~351.41$+$0.64 (2) & 1.30(0.02) & 1.36(0.05) &  1.01(0.03) &   0.90(0.04)\\
G~351.41$+$0.64 (3) & 1.24(0.05) & 1.34(0.08) &  0.90(0.08) &   0.77(0.11)\\
\noalign{\smallskip}\hline\noalign{\smallskip}
\end{tabular}
\label{table:powerlaw}
\end{table}

\normalsize

A comparison of the different power-law indices derived for each individual source
(Table~\ref{table:powerlaw}) shows that
$p$ and $p'$ are nearly the same confirming the idea that optical
depth effects are not important in general.
Mean power-law indices averaged over the sources
in Table~\ref{table:powerlaw}
are: $\langle p\rangle =1.29(0.05)$ and $\langle p'\rangle = 1.34(0.06)$.
We have fitted total integrated intensity maps
without data points with intensities larger than 90\% of the peak,
which could mainly suffer from optical depth effects.
In this case the power-law indices are also close to $p$.
However, a further comparison of power-law indices calculated for total
and reduced maps reveals the following inequality: $p\ge p_{20}\sim p_{40}$
for most of the sources, indicating that N$_2$H$^+$(1--0) intensity
distributions are steeper away from map centers comparing with
inner regions and do not follow single power law.
Note, that the $p_{20}$ and $p_{40}$ indices calculated over the regions
with relatively high intensities could be more influenced by possible
saturation effects than the $p$ indices.
The effect of steepening of dust intensity distributions in outer regions
was found by Beuther et al. (\cite{beuther}) for most of massive star
forming cores they studied; they associated this phenomenon with finite
sizes of the cores.
N$_2$H$^+$(1--0) integrated intensity profiles in low mass cores
also demonstrated the same effect (Paper~I) but it was
explained by fast drop in excitation temperature caused by density
drop in outer core regions.
The latter seems doubtful
since the same cores mapped in CS are usually more extended than
in N$_2$H$^+$ (Zinchenko et al. \cite{zin1}, \cite{zin3}, \cite{zin4}; Juvela \cite{juvela}),
instead, the fast decrease of intensities could be caused
by N$_2$H$^+$ abundance drop (A.~Lapinov, private communication).

Mean power-law indices for reduced maps averaged over the sources
in Table~\ref{table:powerlaw}
are: $\langle p_{20}\rangle =1.05(0.06)$
and $\langle p_{40}\rangle = 0.96(0.10)$.
Assuming excitation conditions are constant and saturation effects are small
in the interior regions where $I/I_{\rm MAX}\ge 0.2$, the derived power-law indices
more likely correspond to density profiles $\propto r^{-2}$
as the theory of isothermal sphere predicts (Shu \cite{shu}).
This is in general consistent with mean power-law index 1.6(0.5)
found by Beuther et al. (\cite{beuther}) for radial density profiles
in inner clump regions ($\le 32\arcsec$) of massive star forming cores
as well as with the value 1.8(0.4) derived by Mueller et al. (\cite{mueller})
for dense cores associated with water masers.
Our sample includes seven sources explored by Mueller et al. (\cite{mueller}),
and for five of them (RNO~1B, NGC~281, G~173.48, S~76E and S~140)
we have got intensity profile power-law indices.
However, density power-law indices derived by Mueller et al. (\cite{mueller})
are systematically lower by $\sim 0.4-0.6$ than the values $p_{20}+1$
expected for density profiles from our data for inner clump regions.
Model calculations of N$_2$H$^+$ excitation in dense cores
as well as maps of optically thin dust emission for the rest of our
core sample are highly desirable to explain the origin of the discrepancy.

\subsection{Velocity dispersion profiles}
\label{sec:vdprof}

The values of line widths obtained from spectra fits are much larger
than thermal widths (kinetic, dust or peak CO(1--0)
temperatures for the sample sources are $\sim 20-50$~K,
see Schreyer et al. \cite{schreyer}, Zinchenko et al. \cite{zin3}, \cite{zin6},
\cite{zin4}) and could be used as a measure for dispersion of non-thermal
velocities in the sources.
This parameter is widely used in theoretical models
describing non-thermal motions of gas within dense cores
and is important for calculations of equation of state,
stability and evolution of dense envelopes around newborn stars.

We have compared line width and integrated intensity maps
and found that in most cases the shapes of the regions with constant line width
do not follow directly integrated intensity distributions
demonstrating more complex clumpy structures.

Since one-dimensional line width profiles demonstrate large scatter,
we used the method from Paper~I to calculate a dependence of averaged
line widths on impact parameters.
Only the values higher than $3\sigma$ were taken for averaging.
Here, the impact parameter ($b$) is the square root
of $A/\pi$, where $A$ is the area of the region enclosing all the points with intensity
larger than certain level, which in turn varies from $I_{\rm max}$
to $0.05\cdot I_{\rm max}$ with a step $0.01\cdot I_{\rm max}$.
If the difference $b_i-b_{i-1}$ becomes higher than half of the mapping
step (10$\arcsec$) the averaged line width and its uncertainty
have been calculated for this region according to the formulas
(\ref{eq:vmean}--\ref{eq:sigmav}).

\begin{figure}
\centering
\includegraphics[width=9cm]{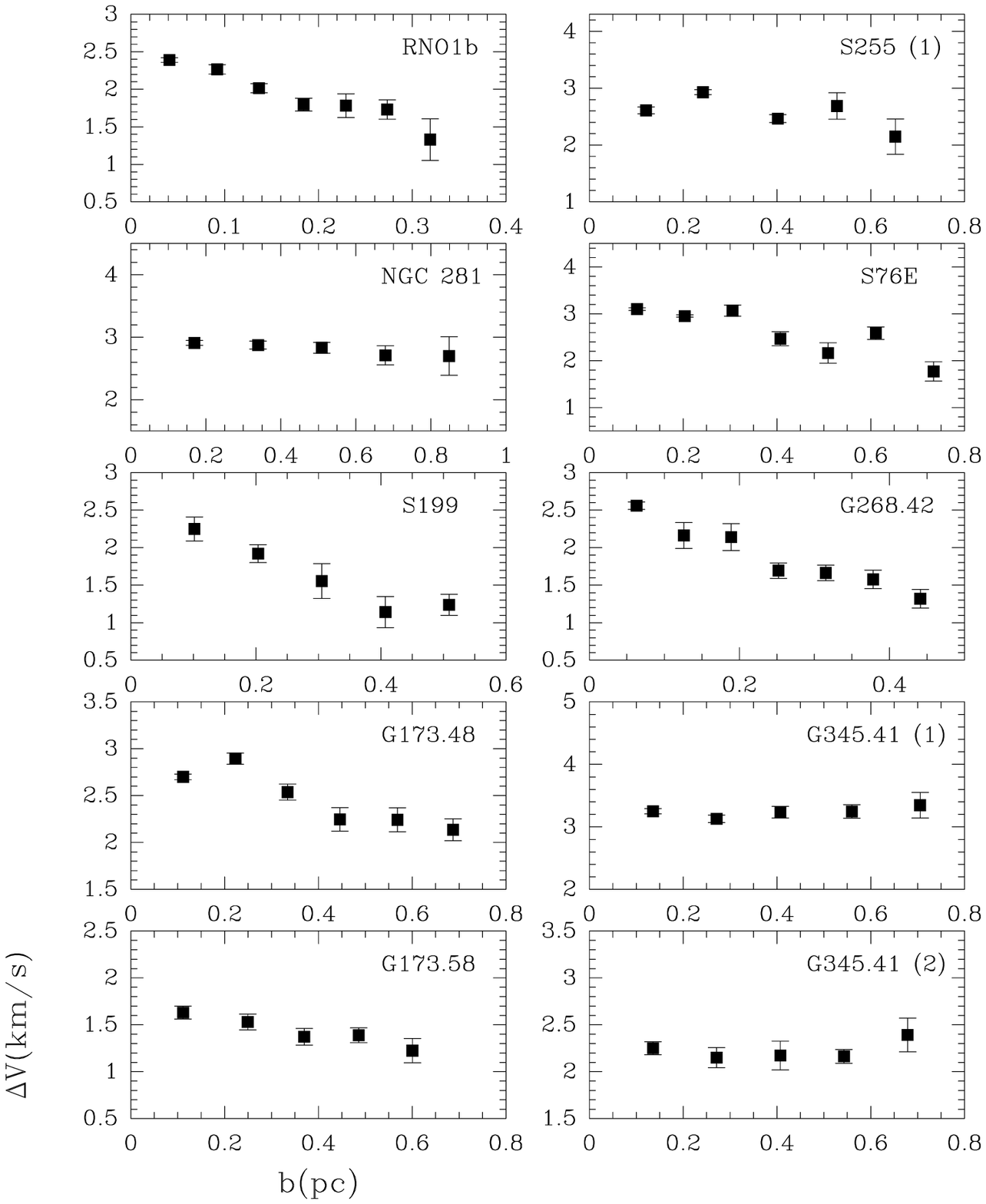}
\caption{Radial velocity dispersion profiles for the cores with one clump
and clumps separated at half maximum intensity level}
\label{vdprof}
\end{figure}

In Fig.~\ref{vdprof} velocity dispersion profiles are given for
ten sources having at least 5 points per profile including
seven cores with one clump and three clumps
separated from their neighbours at half maximum intensity level.
The number of averaged values per point on the plots
increases with $b$ from 1 to 10--15 at the edge.
Uncertainties of averaged line widths also increase with $b$
(if they are lower than $(\sum \sigma_i^{-2})^{-1/2}$,
where $\sigma_i$ is the line width uncertainty in the $i$-th map position
calculated from the spectral fit, the latter values are shown).
The cores with one clump demonstrate that
velocity dispersion decreases with impact parameter
or remains nearly constant in one case (NGC~281).
In 5 of these 7 cases, IRAS sources (taking into account their position
uncertainties) are located within 90\% of peak intensity level.
In the three remaining clumps in Fig.~\ref{vdprof} there is no clear
$\Delta V-b$ trend and either no associated IRAS source or it is located
outside half maximum intensity level.
RNO~1B, S~199 and G~268.42 demonstrate the steepest slopes.
Differences between central line widths and those at the edges of these
cores are $\sim 1-1.5$~km s$^{-1}$.
For eight cores with more than one clump per core and
at least 5 points per profile which are not shown in Fig.~\ref{vdprof}
there is a decrease of line width with impact parameter at least in two cases,
G~269.11 and G~264.28.
In other cases no clear trend has been found.

An enhancement of line width in central regions in several cases
could be at least in part due to optical depth effects.
In fact, line optical depths $\sim 0.5-1.5$ can
result in $\sim 1.1-1.3$ gaussian line broadening due to saturation
(e.g. Phillips et al. \cite{phillips2}).
Therefore, the line width profiles in RNO~1B, S~76E (optical depths
are given in Table~\ref{table:optdepth}) and,
probably, in S~199, G~173.48 and G~173.58, where optical depths
are more uncertain (1.3$\pm$0.7, 0.2$\pm$0.1 and 0.6$\pm$0.3, respectively),
could partially suffer from this effect but cannot alone explain
the observed trends in all the cases.
The conclusion will not change if the objects consist of small
unresolved clumps.

A plausible explanation could be found in a larger degree
of dynamical activity of gas in central regions in the vicinity
of IRAS point sources,
including differential rotation, infall motions and turbulence
due to winds and outflows from massive stars.

\subsection{Velocity gradients}

It is known that an analysis of spatial distributions of line-of-sight
velocities is fruitful in revealing rotational motions of gas
(Goodman et al. \cite{goodman}).
In particular, solid body rotation should produce linear gradients on
$V_{\rm LSR}$ maps.
Velocity gradients also have been interpreted in terms of
turbulent motions by Burkert \& Bodenheimer (\cite{BB}).

In order to investigate velocity structure of the sample cores
we have used a least squares fitting program (Goodman et al.\cite{goodman},
Paper~I) which calculates local and total linear velocity gradients for
a given core (see Paper~I for a detailed explanation of the program).
The results of fits are summarized in Table~\ref{table:vfit} where
the magnitude ($\Gamma$) and the direction ($\Theta_{\Gamma}$, direction of
increasing velocity, measured east of north) of the total velocity gradient
in each core, the product of the velocity gradient and core size
($\Gamma\times d$), and $\beta$, the ratio of rotational to gravitational
energy (Goodman et al. \cite{goodman}, Paper~I) are given.
We have not separated all the cores into clumps
(as has been done in Section~\ref{sec:int-prof}) except those having clumps
that do not overlap at half maximum intensity level
(S~187, G~133.69, AFGL~6366, S~255 and G~345.41).

\begin{table}[htb]
\centering
\caption[]{Results of gradient fitting}
\scriptsize
\begin{tabular}{lrrrr}
\noalign{\hrule}\noalign{\smallskip}
Source   & $\Gamma$               & $\Theta_{\Gamma}$ & $\Gamma\times d$ & $\beta$     \\
         & (km s$^{-1}$ pc$^{-1}$)&   (deg E of N )   &  (km s$^{-1}$)   & ($10^{-3}$) \\
\noalign{\smallskip}\hline\noalign{\smallskip}

RNO~1B              & 1.64(0.04) &   145.8(1.1) & 0.8  & 11.6  \\
NGC~281             & 0.42(0.03) &  --59.3(3.5) & 0.3  & 1.5   \\
S~187 (1)           & 1.10(0.11) &    11.0(6.1) & 0.3  & 16.1  \\
S~187 (2+3)         & 1.47(0.13) & --135.1(4.9) & 0.7  & 28.5  \\
G~133.69+1.22 (1)   & 2.90(0.25) &   141.9(5.3) & 2.3  & 67.8  \\
G~133.69+1.22 (2)   & 0.48(0.09) &  --75.4(5.0) & 0.2  & 2.4   \\
S~199               & 0.27(0.03) &  --24.4(7.4) & 0.2  & 3.6   \\
S~201               & 1.35(0.11) &   143.9(4.0) & 0.6  & 20.0  \\
G~173.48+2.45       & 1.25(0.01) &    26.1(0.8) & 0.8  & 12.1  \\
G~173.58+2.44       & 0.47(0.03) &   149.4(3.1) & 0.4  & 10.8  \\
AFGL~6366 (1)       & 0.28(0.05) &  --89.5(13.3)& 0.1  & 0.4   \\
AFGL~6366 (2)       & 0.99(0.06) & --120.2(3.4) & 0.5  & 10.6  \\
S~255 (1)           & 1.39(0.03) &  --28.9(1.9) & 0.8  & 13.7  \\
S~255 (2)           & 0.59(0.03) &  --35.1(3.4) & 0.4  &  7.4  \\
S~76E               & 0.52(0.01) & --175.9(1.6) & 0.4  & 2.3   \\
DR~21               & 0.41(0.01) &  --95.8(0.4) & 0.5  & 3.9   \\
S~140               & 0.41(0.02) &    58.6(2.5) & 0.2  & 1.4   \\
S~153               & 0.09(0.02) & --122.8(9.5) & 0.2  & 1.0   \\
G~264.28$+$1.48     & 1.15(0.03) &   108.7(2.3) & 0.3  & 3.0   \\
G~265.14$+$1.45     & 0.82(0.01) &  --69.5(0.6) & 0.5  & 7.0   \\
G~268.42$-$0.85     & 0.92(0.03) &   122.7(1.9) & 0.3  & 8.3   \\
G~269.11$-$1.12     & 0.66(0.01) &  --88.4(0.5) & 0.6  & 7.6   \\
G~285.26$-$0.05     & 0.46(0.01) & --123.7(1.3) & 0.7  & 7.2   \\
G 291.27$-$0.71     &0.158(0.004)&   113.3(1.2) & 0.2  & 0.6   \\
G~294.97$-$1.73     & 1.33(0.03) & --112.4(1.2) & 0.5  & 5.1   \\
G~345.41$-$0.94 (1) & 0.81(0.01) & --176.6(1.4) & 0.9  & 10.7  \\
G~345.41$-$0.94 (2) & 1.05(0.02) & --87.3(0.8)  & 1.6  & 70.8  \\
\noalign{\smallskip}\hline\noalign{\smallskip}
\end{tabular}
\label{table:vfit}
\end{table}

\normalsize

In Fig.~\ref{gradients} local velocity gradient maps (white arrows)
overlayed over integrated intensity maps (grey scaled) are shown for 20 cores.
Two objects (S~187 and G~133.69) are not shown due to small number
of points per local gradient map.
The arrows point towards increasing $V_{\rm LSR}$.
Total velocity gradients calculated for the whole cores or distinct clumps
are shown by bold arrows; their values are also indicated.
Many sources show systematic
and at least in some parts of maps
nearly constant (both in magnitude and direction)
velocity gradient fields, implying nearly uniform rotation
(e.g. RNO~1B, S~201, G~173.48, G~173.58, S~255, G~264.28, G~265.14).

\begin{figure}
\centering
\centering \includegraphics[width=9cm]{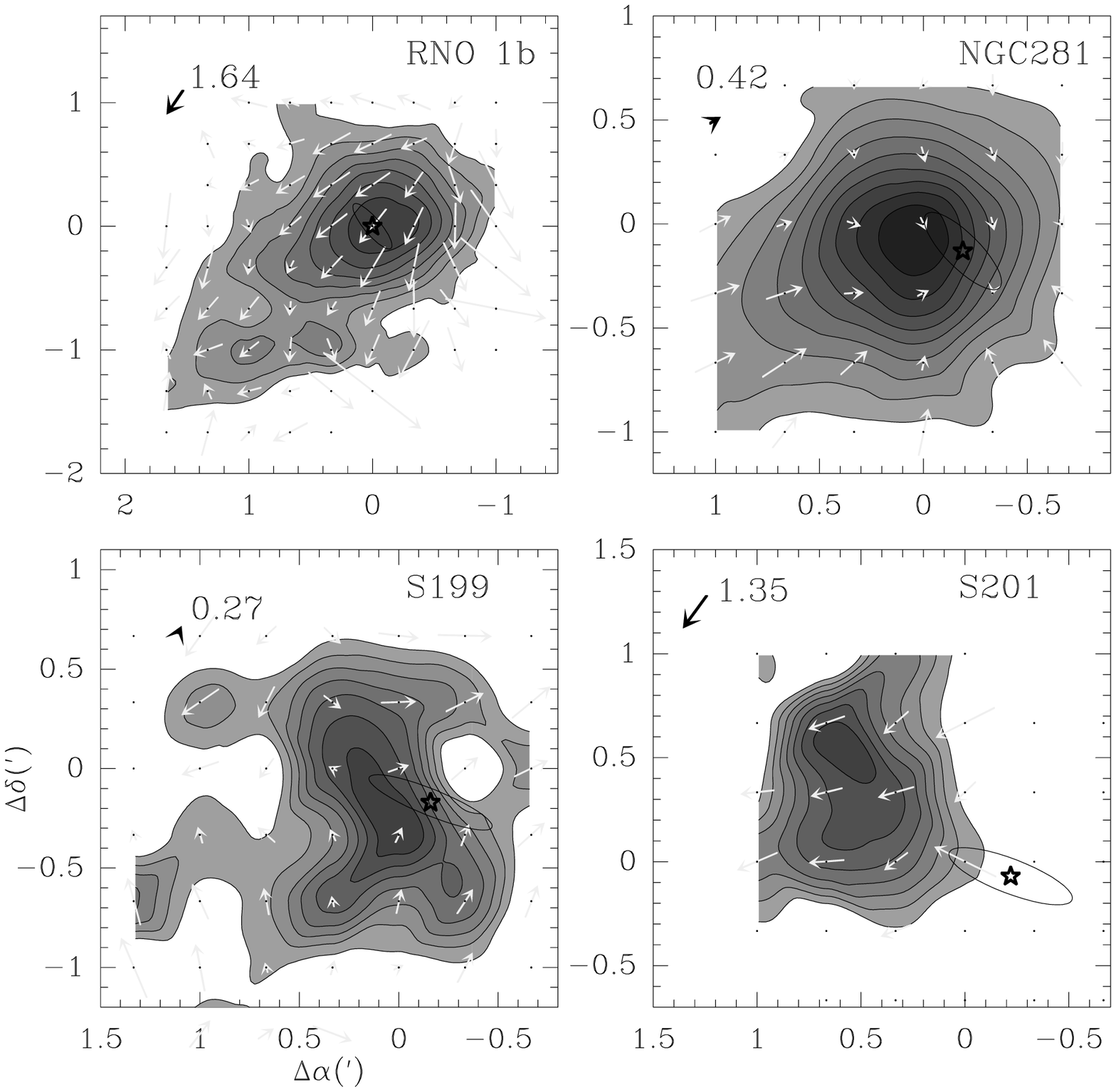}
\caption{Local velocity gradient maps superposed on integrated intensity
maps. Total velocity gradients are shown by bold arrows together with
their values (in km s$^{-1}$ pc$^{-1}$).
IRAS point sources are marked by stars.
The uncertainty ellipses are also shown}
\label{gradients}
\end{figure}

\addtocounter{figure}{-1}

\begin{figure}
\centering
\includegraphics[width=9cm]{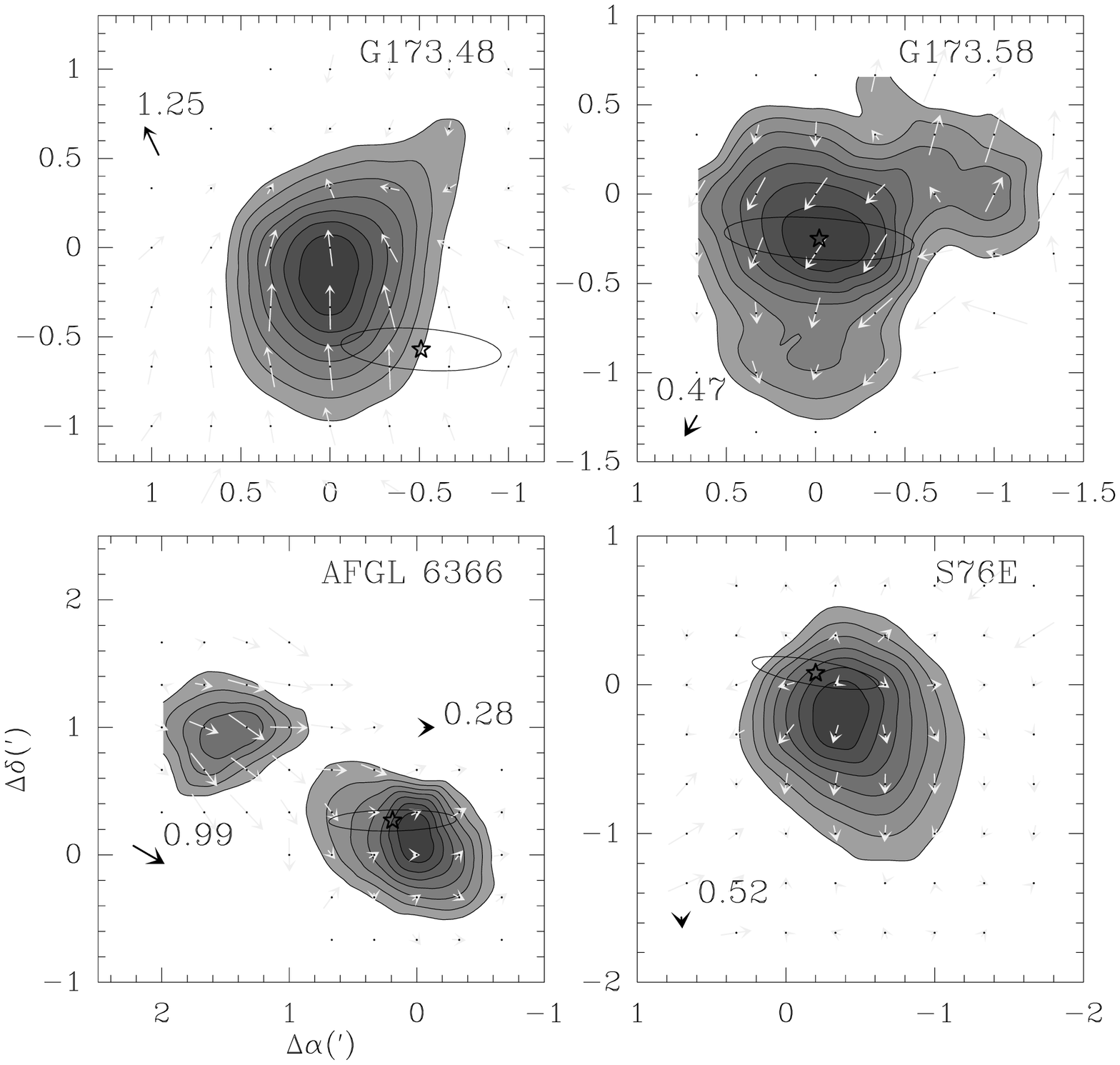}
\caption{Continued}
\end{figure}

\addtocounter{figure}{-1}

\begin{figure}
\centering
\includegraphics[width=9cm]{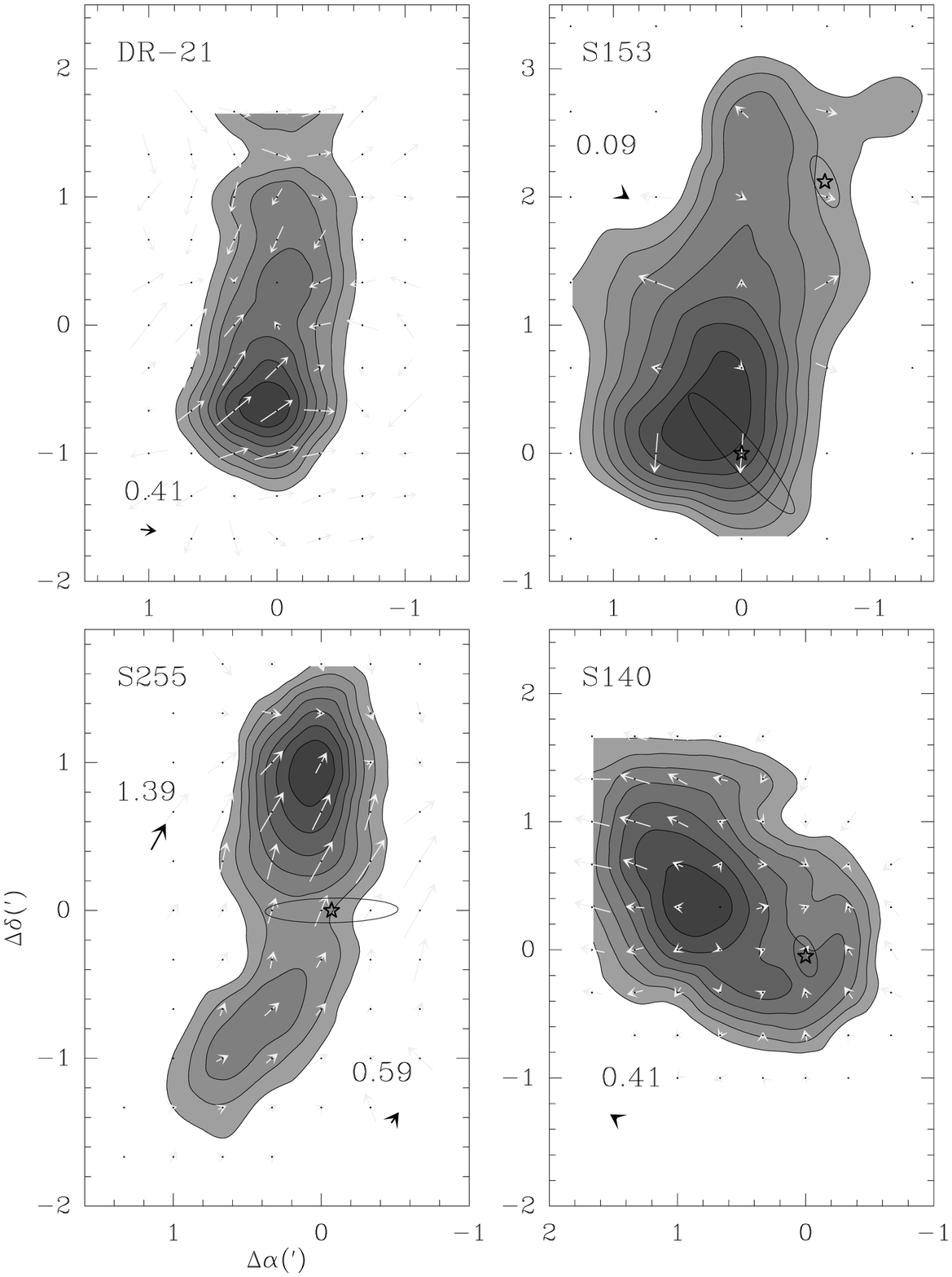}
\caption{Continued}
\end{figure}

\addtocounter{figure}{-1}

\begin{figure}
\centering
\includegraphics[width=9cm]{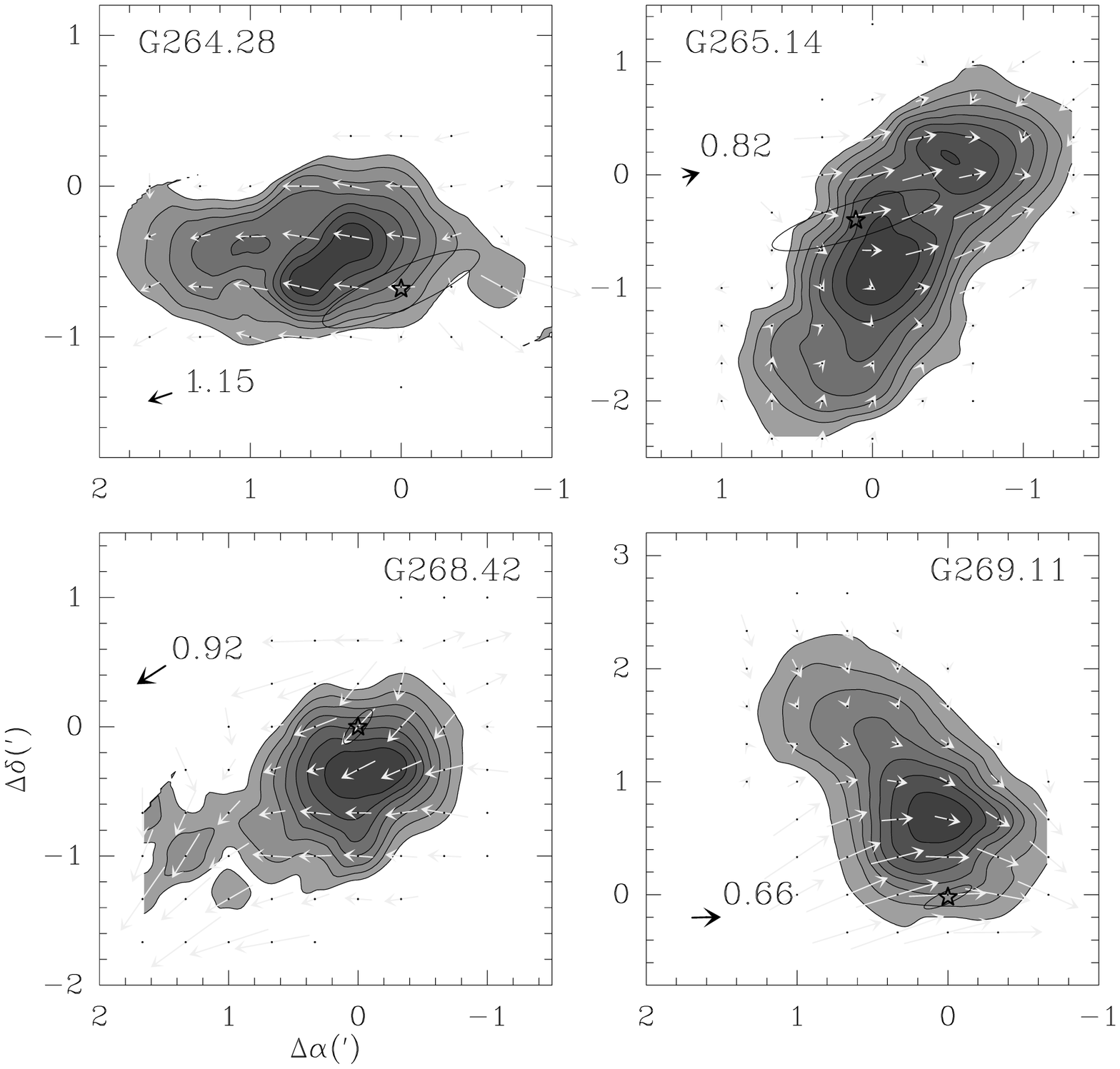}
\caption{Continued}
\end{figure}

\addtocounter{figure}{-1}

\begin{figure}
\centering
\includegraphics[width=9cm]{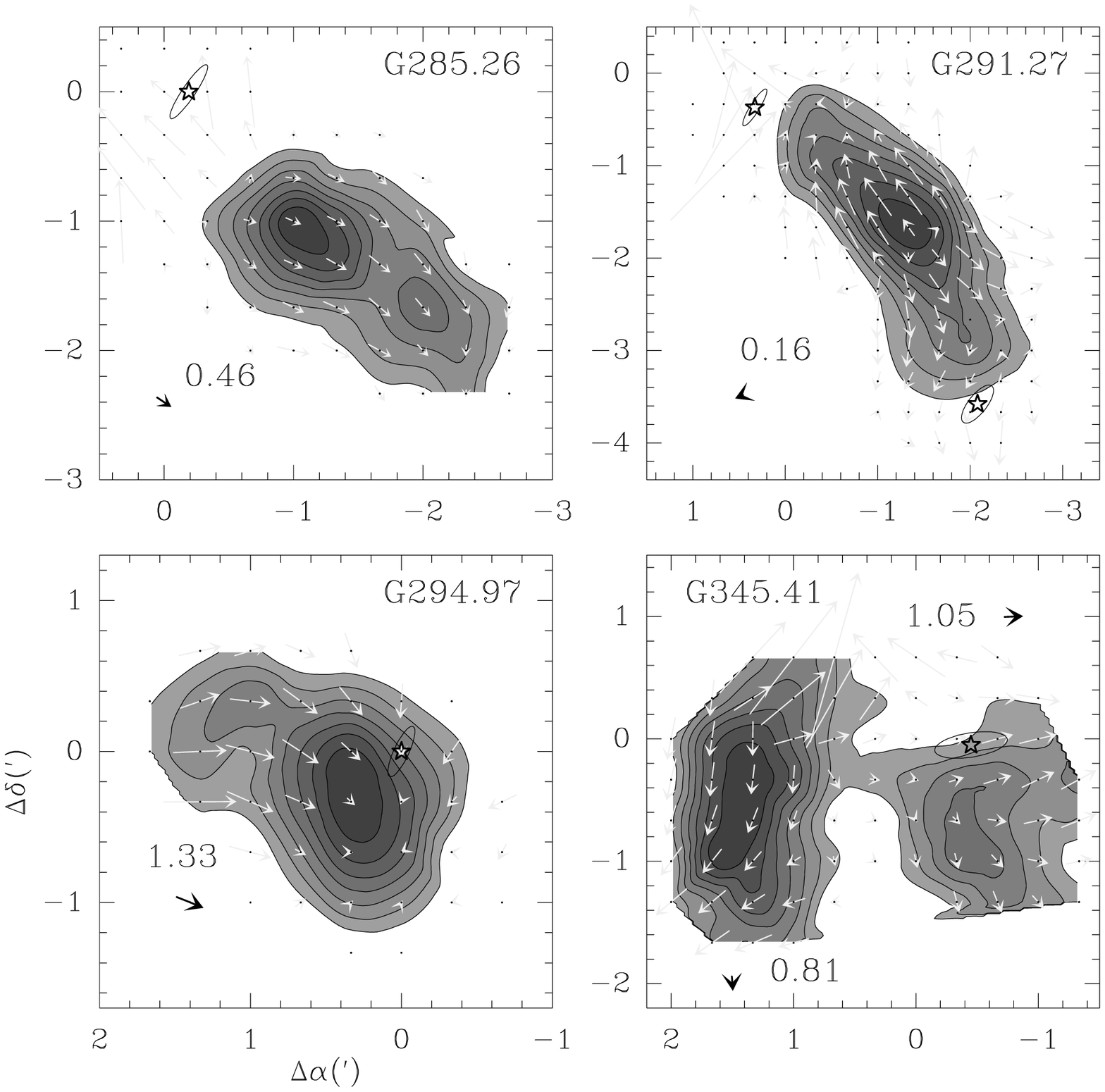}
\caption{Continued}
\end{figure}

The cores with (not separated) closely located clumps
usually have more complex velocity fields than those composed
of single clump or more clumps separated at the half maximum intensity level.
No direct link between velocity gradient distribution
and IRAS source location has been found.
The direction of constant velocity gradient field
in several cases is close to elongation angle of the core.
For 17 cores resolved in both directions
there is a correlation between direction angle
of total velocity gradient ($\Theta_{\Gamma}$) and elongation angle found
from 2D gaussian fitting of integrated intensity maps,
with correlation coefficient, $cc=0.9$.
This fact satisfies an assumption that elongation of cores and clumps
could be due to rotation.
Yet, $\beta$ values vary from $4\cdot 10^{-4}$ to $7.1\cdot 10^{-2}$
with the mean value of 0.01 for 27 sources
(Table~\ref{table:vfit}, last column).
These ratios are the lowest ones among dense cores with
0.2--2~pc sizes (Phillips \cite{phillips}).
Thus, rotation should not play significant role in core dynamics.

\section{Correlations between mean velocity dispersions, column densities
and sizes}

We have searched for possible correlations between mean line widths
(or mean non-thermal velocity dispersions)
and emission region sizes for the cores and clumps
found by deconvolution method (see Section~\ref{sec:deconv}).
Beginning from the works of Larson (\cite{larson}) and Myers (\cite{myers})
such correlations are widely used in estimates of clouds stability
and turbulence.

\begin{figure}
\centering
\includegraphics[width=9cm]{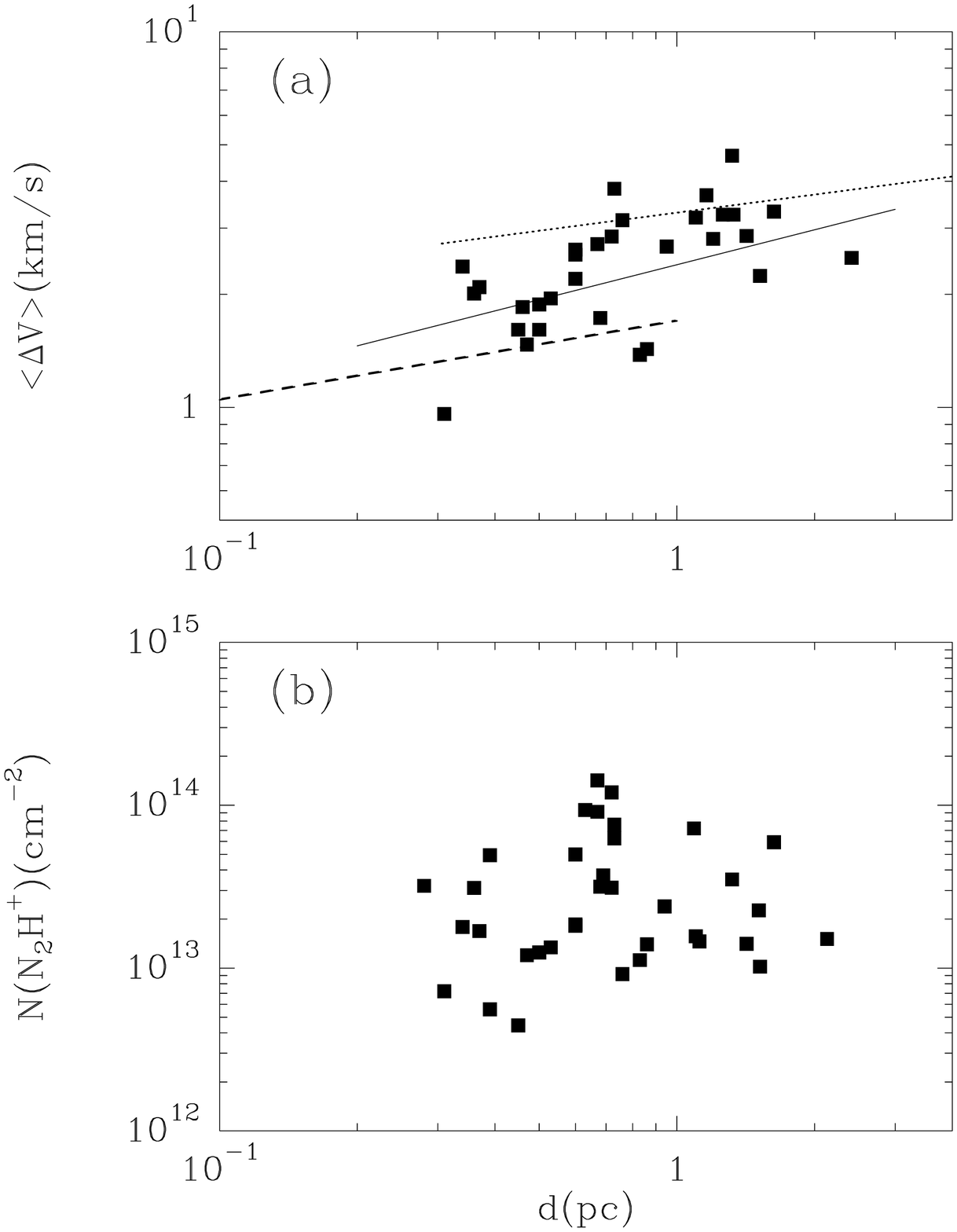}
\caption{Mean line widths versus emission region sizes (a)
together with regression (solid) line.
Dashed and dotted lines correspond to the line width -- size relations
from Caselli \& Myers (\cite{cm}) and Zinchenko (\cite{zin5}), respectively.
Peak N$_2$H$^+$ column densities versus emission region sizes (b)}
\label{corr}
\end{figure}

In Fig.~\ref{corr}(a) we plot mean line widths versus emission
region sizes in logarithmic scale for 26 cores and clumps
having size estimates (Table~\ref{table:phys}), with the exception of three
cores presenting more than one velocity component (G~316.77, G~345.01 and G~351.41).
The correlation coefficient between parameters is not high ($\sim 0.5$),
yet, the correlation is significant.
The hypothesis of zero correlation is rejected at 0.6\% significance level
(Press et al. \cite{press}).
The slope of the regression line, calculated by the ordinary least-squares method,
is 0.3(0.1), and the intercept coefficient is 0.38(0.03).
Practically, the same results have been obtained excluding the cores with
steepest line width profiles (RNO~1B, S~199 and G~268.42,
see Section~\ref{sec:vdprof}), for which mean line widths could be enhanced
due to additional broadening in the central regions of the maps.
However, since the compared variables are independent, it could be more correct
to use linear regression methods that
treat them symmetrically (e.g. orthogonal, reduced major-axis or
bisector methods, see Isobe et al. \cite{isobe}).
Applying these methods to the data we have obtained higher slopes (0.4--0.65)
and nearly the same intercept values.

The regression line calculated by the ordinary least-squares method
is shown in Fig.~\ref{corr}(a) as a solid line together with two additional
linear regressions found by other authors with the same method
for massive cores.
Whereas the dependence of Caselli \& Myers (\cite{cm}) (dashed line)
was calculated using multi-line data for more compact cores in Orion A and B,
the dependence of Zinchenko (\cite{zin5}) (dotted line) was calculated
for the sample observed in CS and C$^{34}$S.
All the slopes agree within the uncertainties of their determination.

We have also searched for possible correlations between
peak N$_2$H$^+$ column densities (which in the case of constant
N$_2$H$^+$ abundances trace hydrogen column densities) and sizes
(Fig.~\ref{corr}(b)).
These parameters have been chosen because they
are independent, unlike to mean densities and virial masses
which have been calculated using size estimates (see Section~\ref{sec:phys}).
No significant correlation has been found.
No correlation has been also found between mean N$_2$H$^+$ column densities
(averaged over emission regions with $I/I_{\rm MAX}\ge 0.5$)
and sizes.

It is doubtful that the observed $\langle\Delta V\rangle-d$ dependence is
connected with general turbulence properties
or with some common equation of state for the sample cores.
As it is shown in Section~\ref{sec:vdprof} there is no evidence
for growing line width with distance inside the cores.
On the other hand, the dependence found does not contradict
the assumption of gravitationally bounded objects in virial equilibrium.
Depending on the so-called virial parameter, $M_{\rm vir}/M$
(Bertoldi \& McKee \cite{BM}), where virial mass
is calculated from equation (\ref{eq:mvir}) and $M$ is an independent estimate of mass,
such objects could be in the critical state on the threshold of gravitational collapse
($M_{\rm vir}/M\sim$1) or in pressure equilibrium ($M_{\rm vir}/M>>$1)
(e.g. Pirogov \& Zinchenko \cite{pz}).
If the cores are close to the critical state, with magnetic energy being
of the order of kinetic energy, the slope of
the $\langle\Delta V\rangle-d$ dependence should be equal to 0.5, with
intercept coefficient depending on external pressure and inner density profile.
Column density in this case does not depend on size.
If the cores are in pressure equilibrium, any
$\langle\Delta V\rangle\propto d^{\,q}$
dependence implies a corresponding $M\propto d^{\,3-2q}$ dependence.

Our virial masses for four cores coincide within a factor of 1.5 with
the $M_n$ masses found by Mueller et al. (\cite{mueller}), and
in one case (RNO~1B) the virial mass is $\sim 5$ times lower
than the corresponding $M_n$ value,
which could be connected with differences in core size.
Masses derived from CS observations for most of the sample sources
(Zinchenko et al. \cite{zin1}, \cite{zin3}, \cite{zin4}; Juvela \cite{juvela})
are several times larger than our virial masses (except G~316.77), yet,
CS emission is usually more extended.
Therefore, if the studied cores are in equilibrium, they could be close
to the critical state rather than in pressure equilibrium.

\begin{table*}[htb]
\centering
\caption[]{Physical parameters of massive and low-mass cores}
\begin{tabular}{lcccccc}
\noalign{\hrule}\noalign{\smallskip}
Parameter     & \multicolumn{2}{c}{Massive cores} & \multicolumn{4}{c}{Low-mass cores} \\
              &                &                  & \multicolumn{2}{c}{With stars} & \multicolumn{2}{c}{Without stars} \\
              & Mean           & Number           & Mean       & Number            & Mean     & Number \\
\noalign{\smallskip}\hline\noalign{\smallskip}
$\Delta V$ (km s$^{-1}$)    & 2.5(0.8)&  31    & 0.5(0.3)   &  35    & 0.3(0.1) & 25  \\
$N$(N$_2$H$^+$)
($10^{12}$ cm$^{-2}$)& 29(33)   &  36    &  8(5)      &  35    & 6(3)     & 25  \\
$d$(pc)              & 0.5(0.2) &  36    & 0.14(0.03) &  22    &0.1(0.02) & 13  \\
Axial ratio          & 1.4(0.4) &  47    &  1.9(0.7)  &  35    & 2(1)     & 19  \\
$n_{\rm vir}$
($10^4$ cm$^{-3}$)   &  6(4)    &  36    & 20(20)     &  34    & 12(7)    & 19  \\
$M_{\rm vir}$
($M_{\odot}$)        & 658(604) &  36    & 9(16)      &  34    & 3(2)     & 19  \\
$X$(N$_2$H$^+$)
($10^{-10}$)         &   5(3)   &  36    & 3(2)       &  34    & 2(1)     & 18  \\
$p$                  & 1.3(0.1) &  25    & 1.2(0.1)   &  9     & 0.8(0.04)& 8   \\
$\Gamma$
(km s$^{-1}$ pc$^{-1}$) & 0.5(0.4)&  27    & 2(1)       &  14    & 2(1)     & 12  \\
$\beta$              & 0.01(0.02) &  27    & 0.02(0.02) &  13    &0.02(0.02)& 7   \\
\noalign{\smallskip}\hline\noalign{\smallskip}
\end{tabular}
\label{table:comparison}
\end{table*}

\section{Comparison with physical parameters of low-mass cores}

In Table~\ref{table:comparison} physical parameters of the clumps
averaged over the studied sample are given together with the same parameters
derived in Paper~I for low-mass cores with and without stars.
Physical parameters for massive cores are taken from Table~\ref{table:line}
and Table~\ref{table:phys}.
We have used weighted averages for the parameters which uncertainties
resulted from fits (line width, size, axial ratio, power-law index,
velocity gradient),
for the others standard arithmetic means are given.
In brackets we give standard deviations (as in Paper~I).

Apart from well-known differences in virial mass and velocity dispersion
between these two samples,
one can see that massive cores have higher sizes and column densities,
lower mean densities and velocity gradients than low-mass counterparts.
Note, that for low-mass cores $d=2r$, where $r$ was taken
from Table~6 of Paper~I.
Mean power-law indices for integrated intensity
maps for the studied cores are also given
as well as the same parameters for low-mass cores
(see Table~6 from Paper~I, $p=\alpha-1$).
On average, integrated intensity profiles in massive cores have nearly the same
power-law indices as in low-mass cores within uncertainties of their
determination.
Mean axial and $\beta$ ratios, as well as mean N$_2$H$^+$ abundances
also are nearly the same taking into account their uncertainties.
Velocity dispersion in massive cores is either constant or decreases
with distance from the center in contrast to low-mass cores
where no common tendency was found.

The fact that mean density in massive cores is lower
than in low-mass cores is somewhat unexpected.
For example, Fontani et al. (\cite{fontani}) recently found densities of
$10^5-10^6$~cm$^{-3}$ for the sample of massive cores associated with
ultracompact H~II regions and having sizes 0.2--1.6~pc.
Possible explanation could be found if dense cores from our sample
actually consist of unresolved clumps much smaller than the telescope beam.
If density of such unresolved clumps is the same as for
low-mass cores with stars (see Table~\ref{table:comparison})
the clumps volume filling factor should be $\sim 0.3$.

\section{Conclusions}

In order to derive physical conditions of dense gas in
molecular cloud cores where massive stars and star clusters are formed,
35 objects from northern and southern hemispheres have been observed
in the N$_2$H$^+$(1--0) line.
N$_2$H$^+$ emission has been detected in 33 sources, and detailed maps
have been obtained for 28 sources.
The results can be summarized as follows:

\begin{enumerate}
\item
Peak LTE N$_2$H$^+$ column densities lie in the range:
3.6$\cdot$10$^{12}$--1.5$\cdot$10$^{14}$~cm$^{-2}$.
Intensity ratios of (01--12) to (23--12) hyperfine components
in most cases are slightly higher than the LTE value.
The optical depth of (23--12) component toward peak intensity positions
of 10 sources is $\sim 0.2-1$.
\item
N$_2$H$^+$ intensity maps in most cases show elongated or more
complex structures.
Two-dimensional gaussian fittings have revealed
47 clumps in 26 objects.
Their sizes lie in the range:
0.3--2.1~pc, the range of virial masses is $\sim 30-3000~M_{\odot}$.
Mean N$_2$H$^+$ abundance for 36 clumps is $(5.2\pm 0.5)\cdot 10^{-10}$.
\item
Power-law indices for radial integrated intensity profiles in the cores
with axial ratios $<2$ have been calculated.
The mean value of power-law index is close to 1.3 for total integrated
intensity maps in 25 objects,
and about unity for reduced maps, where positions of low intensity were
rejected.
This corresponds to a density profile $\propto r^{-2}$ in the core inner regions,
assuming constant N$_2$H$^+$ excitation conditions and abundances,
and small saturation effects.
\item
In the cores with relatively extensive and high quality maps
N$_2$H$^+$ line widths either decrease or stay constant
with distance from the center implying an enhanced dynamical activity
in the central regions in the vicinity of IRAS sources.
\item
Elongation angles of the 17 cores and distinct clumps correlate
with total velocity gradient direction confirming that elongation could be
due to rotation.
However, the ratio of rotational to gravitational energy
($4\cdot 10^{-4}$--$7.1\cdot 10^{-2}$)
is too low for rotation to play a significant role in the dynamics of the cores.
\item
A correlation has been found between mean line widths and sizes of the clumps,
$\langle \Delta V \rangle\propto d^{\,0.3\pm 0.1}$.
No correlation has been found between peak N$_2$H$^+$ column densities
and the sizes.
These results do not contradict the assumption of gravitationally
bounded objects in virial equilibrium.
\item{}
A comparison between mean physical parameters of massive
and low-mass cores (Paper~I) reveals that massive cores have higher sizes,
N$_2$H$^+$ column densities, virial masses and velocity dispersions
as well as lower mean densities and velocity gradients than low-mass
counterparts.
Mean N$_2$H$^+$ abundances, axial ratios and ratios of rotational
to gravitational energy are nearly the same.
N$_2$H$^+$ integrated intensity profiles have nearly the same mean
power-law indices while velocity dispersions in massive cores
have a tendency to decrease with distance
in contrast to low-mass cores, where no common trend was found.

\end{enumerate}

\begin{acknowledgements}

We are grateful to A.~Lapinov for valuable comments.
We would like to thank the anonymous referee for useful detailed comments.
The research has made use of the SIMBAD database,
operated by CDS, Strasbourg, France.
The work was supported by  NASA-CRDF grant RPO-841,
INTAS grant 99-1667 and Russian Foundation for Basic
Research grant 03-02-16307 (in part).
PC wish to acknowledge support from the MURST project
``Dust and molecules in astrophysical environments".
PCM would like to acknowledge NASA grant NAG-6266.

\end{acknowledgements}

{}

\end{document}